\documentclass[twocolumn,showpacs,aps,prd,floatfix,preprintnumbers]{revtex4}
\usepackage{graphicx} 
\usepackage{dcolumn}
\usepackage{epsfig}    
\usepackage{amsmath}
\usepackage{color}

\input babarsym   

\renewcommand{\kk}{\ensuremath{\Kp\Km}\xspace}
\renewcommand{\pipi}{\ensuremath{\pip\pim}\xspace}
\newcommand{\gkk}{\ensuremath{\gamma\Kp\Km}\xspace}
\newcommand{\gpipi}{\ensuremath{\gamma\pip\pim}\xspace}
\newcommand{\mmtwo}{\ensuremath{M^2_{\mathrm{rec}}}\xspace}
\renewcommand{\mm}{\ensuremath{M_{\mathrm{rec}}}\xspace}

\newcommand{\pips}{\ensuremath{\pi_s^+}\xspace}
\newcommand{\pims}{\ensuremath{\pi_s^-}\xspace}

\def\etal {{\it et al.}}
\def\pis {{\ensuremath{\pi_s}\xspace}}
\mathchardef\myhyphen="2D
\newcommand{\BaBarPubYear}    {18}
\newcommand{\BaBarPubNumber}  {003}
\newcommand{\SLACPubNumber}   {17246}
\newcommand{\al}{\ensuremath{\kern 0.5em }}
\newcommand{\all}{\ensuremath{\kern 0.25em }}

\begin{document}
\begin{flushleft}
\babar-PUB-\BaBarPubYear/\BaBarPubNumber \\
SLAC-PUB-\SLACPubNumber \\
\end{flushleft}
\title{
 \large \bf\boldmath Study of \OneS Radiative Decays to \gpipi and \gkk
}

%
\author{J.~P.~Lees}
\author{V.~Poireau}
\author{V.~Tisserand}
\affiliation{Laboratoire d'Annecy-le-Vieux de Physique des Particules (LAPP), Universit\'e de Savoie, CNRS/IN2P3,  F-74941 Annecy-Le-Vieux, France}
\author{E.~Grauges}
\affiliation{Universitat de Barcelona, Facultat de Fisica, Departament ECM, E-08028 Barcelona, Spain }
\author{F.~Giannuzzi$^{a}$}
\author{A.~Palano$^{ab}$}
\affiliation{INFN Sezione di Bari$^{a}$; Dipartimento di Fisica, Universit\`a di Bari$^{b}$, I-70126 Bari, Italy }
\author{G.~Eigen}
\affiliation{University of Bergen, Institute of Physics, N-5007 Bergen, Norway }
\author{D.~N.~Brown}
\author{Yu.~G.~Kolomensky}
\affiliation{Lawrence Berkeley National Laboratory and University of California, Berkeley, California 94720, USA }
\author{M.~Fritsch}
\author{H.~Koch}
\author{T.~Schroeder}
\affiliation{Ruhr Universit\"at Bochum, Institut f\"ur Experimentalphysik 1, D-44780 Bochum, Germany }
% ...
\collaboration{The \babar\ Collaboration}
\noaffiliation

\begin{abstract}
We study the \OneS radiative decays to \gpipi and \gkk using data 
recorded with the \babar\ detector operating at the SLAC PEP-II
asymmetric-energy \epem\ collider at center-of-mass energies at the
\TwoS and \ThreeS resonances.
The \OneS resonance is reconstructed from the decay 
$\Upsilon(nS)\to \pi^+ \pi^- \OneS$, $n=2,3$.
Branching fraction measurements and spin-parity analyses of \OneS
radiative decays are reported 
for the I=0 $S$-wave and $f_2(1270)$ resonances in the $\pip \pim$ mass
spectrum, 
the $f_2'(1525)$ and $f_0(1500)$ in the \kk mass spectrum, and the $f_0(1710)$ in both.
\end{abstract}
\maketitle

  \section{Introduction}
  
The existence of gluonium states is still an open issue for
Quantum Chromodynamics (QCD). Lattice QCD calculations predict the lightest gluonium states
to have quantum numbers $J^{PC}=0^{++}$ and $2^{++}$ and to be in
the mass region below 2.5 \gevcc~\cite{lattice}.
In particular, the $J^{PC}=0^{++}$ glueball is predicted to have a mass around 1.7 \gevcc.
Searches for these
states have been performed using many supposed ``gluon rich''
reactions. However, despite intense experimental searches, there
is no conclusive experimental evidence for their direct observation~\cite{klempt,ochs}.
The identification of the scalar glueball is further complicated by the possible mixing
with standard $q \bar q$ states.
The broad $f_0(500)$, $f_0(1370)$~\cite{mink}, $f_0(1500)$~\cite{amsler1,amsler2}, and 
$f_0(1710)$~\cite{gg} have been suggested as scalar glueball candidates.
A feature of the scalar glueball is that its $s \bar s$ decay mode should be favored with
respect to 
$u \bar u$ or $d \bar d$ decay modes~\cite{chano,chao}.

Radiative decays of heavy quarkonia, in which a photon replaces one of the three gluons
from the strong decay of \jpsi or \OneS, can probe color-singlet
two-gluon systems that produce gluonic resonances. 
Recently, detailed calculations have been performed on the production
rates of the scalar glueball in the process $V(1^{--}) \to \gamma G$,
where $G$ indicates the scalar glueball and $V(1^{--}) $ indicates
charmonium or bottomonium vector
mesons such as $J/\psi$, $\psi(2S)$, or
\OneS~\cite{zhu,fleming1,fleming2, he}.

\jpsi decays
have been extensively studied in the past~\cite{kopke} and are
currently analyzed in $\ep \en$ interactions by BES experiments~\cite{bes1,bes2}.
The experimental observation of radiative \OneS decays is challenging because their rate
is suppressed by a factor of $\approx 0.025$ compared
to \jpsi radiative decays, which are of order $10^{-3}$~\cite{cleo1}.
Radiative \OneS decays to a pair of hadrons have been studied by the CLEO collaboration~\cite{cleo1,cleo2} with limited
statistics and large backgrounds from $\epem \to \gamma \
(vector \ meson)$.
In this work we observe \OneS decays through the decay chain
$\TwoS/\ThreeS \to \pip \pim \OneS$. This allows us to study \OneS
radiative decays to $\pip \pim$ and $\Kp \Km$ final states with 
comparable statistics, but lower background.

This paper is organized as follows. In Sec. II, we give a brief description of the
\babar\ detector and Sec. III is devoted to the description of event reconstruction.
In Sec. IV we study resonance production in $\pip \pim$ and $\Kp \Km$ final states and Sec. V is devoted 
to the description of the efficiency correction. We describe in Sec. VI a study of the angular distributions
using a Legendre polynomial moments analysis while Sec. VII gives results on the full angular analysis.
The measurement of the branching fractions is described in Sec. VIII and the results are summarized in Sec. IX.

\section{The \babar\ detector and dataset}

The results presented here are based on data collected
by the \babar\ detector 
with the PEP-II asymmetric-energy $e^+e^-$ collider
located at SLAC,  at the \TwoS and \ThreeS resonances with integrated
luminosities~\cite{lumy}
of 13.6 and 28.0 \invfb, respectively.
The \babar\ detector is described in detail elsewhere~\cite{BABARNIM}.
The momenta of charged particles are measured
by means of a five-layer, double-sided microstrip detector,
and a 40-layer drift chamber, both operating  in the 1.5~T magnetic 
field of a superconducting solenoid. 
Photons are measured and electrons are identified in a CsI(Tl) crystal
electromagnetic calorimeter (EMC). Charged-particle
identification is provided by the measurement of specific energy loss in
the tracking devices, and by an internally reflecting, ring-imaging
Cherenkov detector. 
Muons and \KL\ mesons are detected in the
instrumented flux return  of the magnet.
Monte Carlo (MC) simulated events~\cite{geant}, with reconstructed sample sizes 
more than 100  times larger than the corresponding data samples, are
used to evaluate the signal efficiency.

\section{Events reconstruction}

We reconstruct the decay chains
\begin{equation}
  \TwoS/\ThreeS \to (\pips \pims)  \OneS \to (\pips \pims) (\gamma \pip \pim) 
\label{eq:twoa}
\end{equation}
and
\begin{equation}
  \TwoS/\ThreeS \to (\pips \pims)  \OneS  \to (\pips \pims) (\gamma \Kp \Km),
\label{eq:twob}
\end{equation}
where we label with the subscript $s$ the slow pions from the
direct \TwoS and \ThreeS decays.
We consider only events containing exactly four well-measured
tracks with transverse momentum greater than 0.1~\gevc\ and a total net
charge equal to zero.
We also require exactly one well-reconstructed $\gamma$ in
the EMC having an energy greater than 2.5 \gev. To remove background
originating from \piz mesons we remove events having
\piz\ candidates formed with photons having an energy greater than
100 \mev.
The four tracks 
are fitted to a common vertex, with the requirements that the fitted vertex be within the
$e^+ e^-$ interaction region and have a $\chi^2$ fit probability greater than 0.001.
We select muons, electrons, kaons, and pions by applying high-efficiency particle identification
criteria~\cite{book}. For each track we test the
electron and muon identification hypotheses and remove the event if
any of the charged tracks satisfies a tight muon or electron
identification criterion.

We require momentum balance for the four final states, making use of
a $\chi^2$ distribution defined as
\begin{equation}
\chi^2 = \sum_{i=1}^{3}\frac{(\Delta {\bf p}_i - \langle\Delta {\bf p}_i\rangle)^2}{\sigma_i^2},
\label{eq:chi}
\end{equation}
where $\Delta {\bf p}_i $ are the missing laboratory three-momenta components
\begin{equation}
\Delta {\bf p}_i = {\bf p}_i^{e^+} + {\bf p}_i^{e^-} - \sum_{j=1}^{5}{\bf p}_i^j,
\label{eq:chi1}
\end{equation}
and $\langle\Delta {\bf p}_i \rangle$ and $\sigma_i$ are the mean values and the widths of
the missing momentum distributions. These are obtained from signal MC
simulations of the four final states through two or
three Gaussian function fits to the MC balanced momentum
distributions. 
When multiple Gaussian functions are used, the mean values and
$\sigma$ quoted are
average values weighted by the relative fractions. 
In Eq.
(\ref{eq:chi1}), ${\bf p}_i$
indicates the three components of the
laboratory momenta of the five particles in the final state, while
${\bf p}_i^{e^+}$
and ${\bf p}_i^{e^-}$ indicate the three-momenta of the incident beams.

\begin{figure}
\begin{center}
\includegraphics[width=8.5cm]{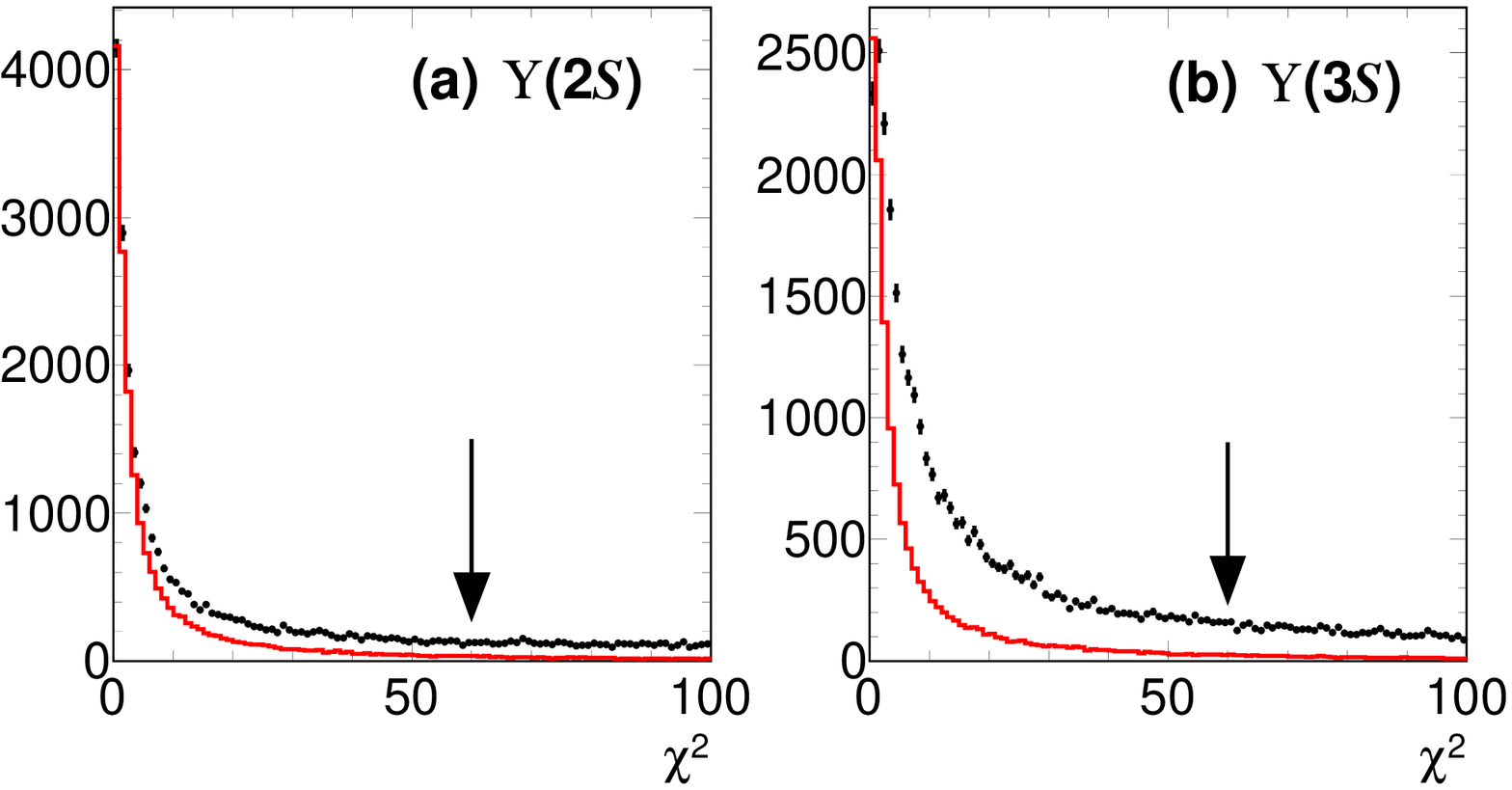}
\caption{$\chi^2$ distributions used for defining the momentum balance
 for data (black dots) compared with signal MC simulations (full (red) line) for
 (a) \TwoS and (b) \ThreeS data from reactions~(\ref{eq:twoa}). 
The arrows indicate the cut-off used to select momentum balancing events.}
\label{fig:fig1}
\end{center}
\end{figure}

Figure~\ref{fig:fig1} shows the $\chi^2$ distributions for reactions
(a) $\TwoS \to (\pips \pims)  \OneS \to (\pips \pims) (\gamma \pip \pim)$ and
(b) $\ThreeS \to (\pips \pims)  \OneS \to (\pips \pims) (\gamma \pip \pim)$, respectively
compared with signal MC simulations. 
The accumulations at thresholds represent events satisfying momentum
balance. We apply a very loose selection,
$\chi^2<60$, optimized using the \TwoS data, and remove events
consistent with being entirely due to background. 
We note a higher background in the \ThreeS data, but keep the same loose 
selection to achieve a similar  efficiency.

Events with balanced momentum are then required to satisfy energy balance
requirements. In the above decays the $\pis$ originating from 
direct \TwoS/\ThreeS decays have a 
soft laboratory momentum distribution ($<600 \mevc$), partially overlapping with the
hard momentum distributions for the hadrons originating from the \OneS
decay.
We therefore require energy balance, following a combinatorial approach. 

\begin{figure}
\begin{center}
\includegraphics[width=8.5cm]{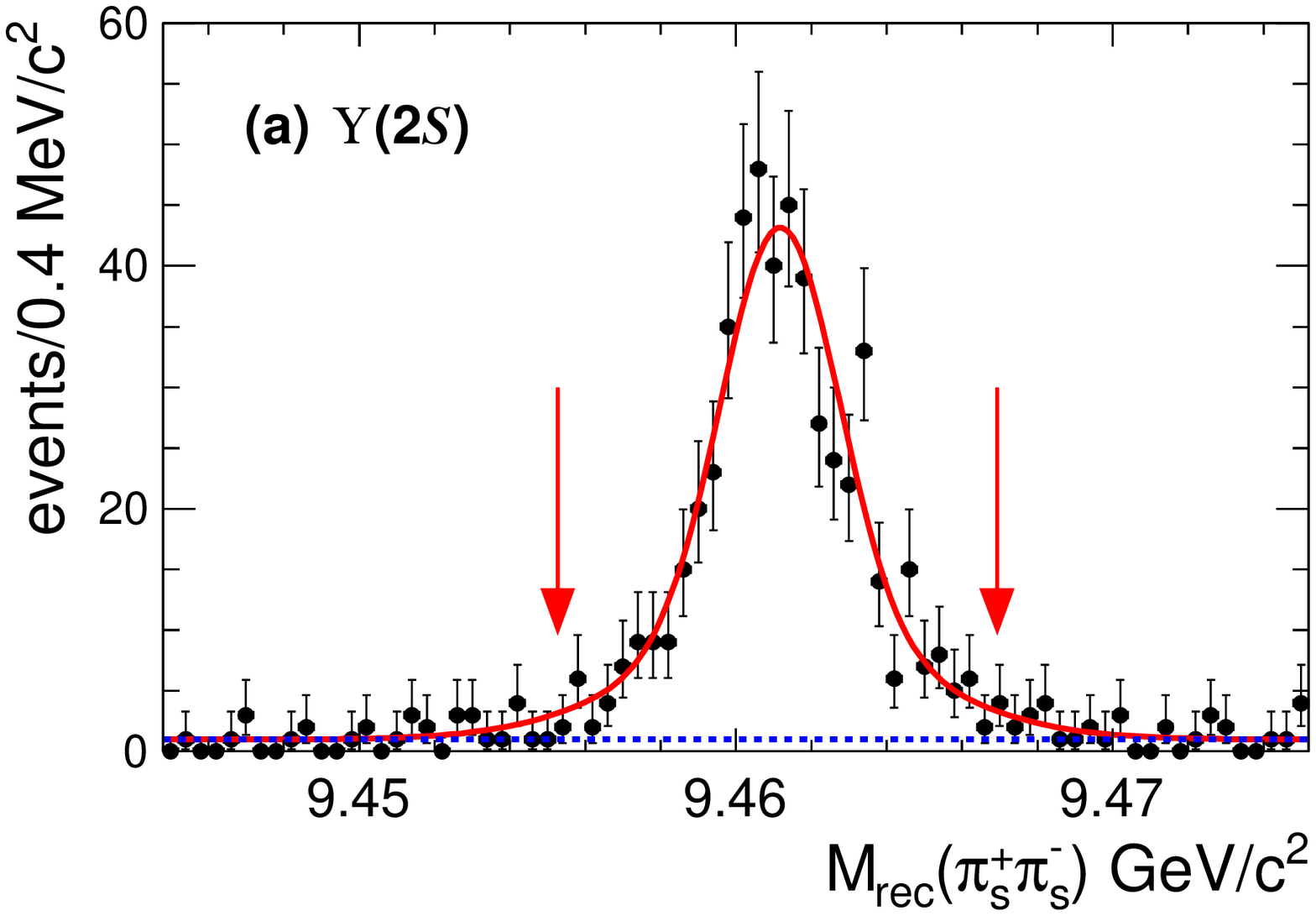}
\includegraphics[width=8.5cm]{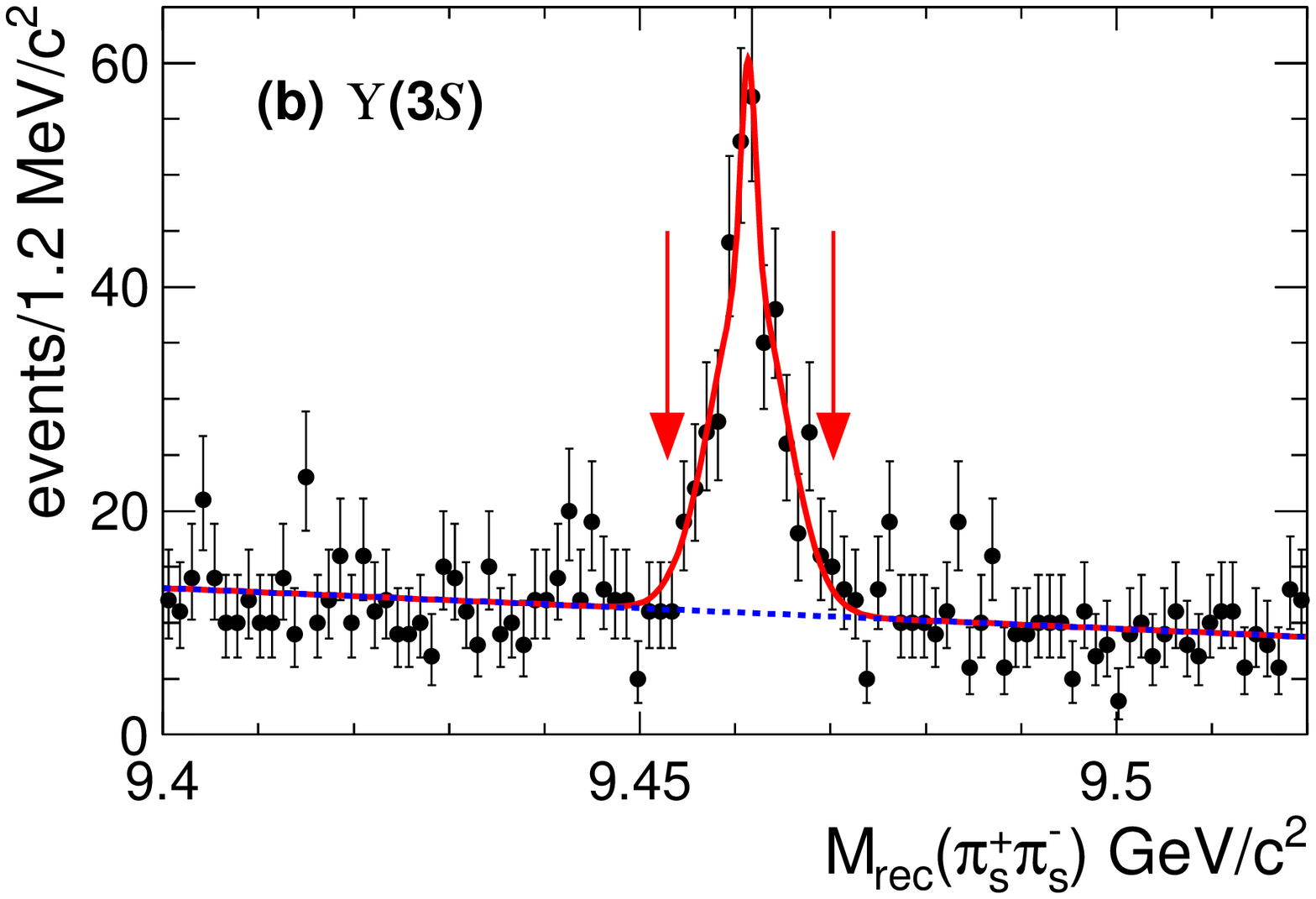}
\caption{Combinatorial recoiling mass \mm to $\pips
  \pims$ candidates for
 (a) \TwoS and (b) \ThreeS data. The lines are the results from the
 fit described in the text. The arrows indicate the selections used to 
apply the energy balance criterion.}
\label{fig:fig2}
\end{center}
\end{figure}

For
each combination of $\pips \pims$ candidates, we first require both
particles to be identified loosely as pions and compute the recoiling
mass
\begin{equation}
\mmtwo(\pips \pims) = |p_{\ep} + p_{\en} - p_{\pips} - p_{\pims}|^2,
\end{equation}
where $p$  is the particle four-momentum. The distribution
of $\mmtwo(\pips \pims)$ is expected to peak at the squared \OneS mass for signal events.
Figure~\ref{fig:fig2} shows the combinatorial recoiling mass $\mm(\pips \pims)$
for $\TwoS$ and $\ThreeS$ data, where narrow peaks at the \OneS mass
can be observed.

We fit each of these distributions using a linear function for the background
and the sum of two Gaussian functions for the signal, obtaining average $\sigma=2.3$ \mevcc\ and $\sigma=3.5$ \mevcc\ values
for the \TwoS and \ThreeS data, respectively. We select signal event
candidates by requiring
\begin{equation}
|\mm(\pi^+_s \pi^-_s) - m(\OneS)_f|<2.5 \sigma ,
\label{eq:mrec}
\end{equation}
where $m(\OneS)_f$ indicates the fitted \OneS mass value. We obtain, in the above mass window, values of
signal-to-background ratios of 517/40 and 276/150 for \TwoS and \ThreeS data, respectively.

\begin{figure}
\begin{center}
\includegraphics[width=8.5cm]{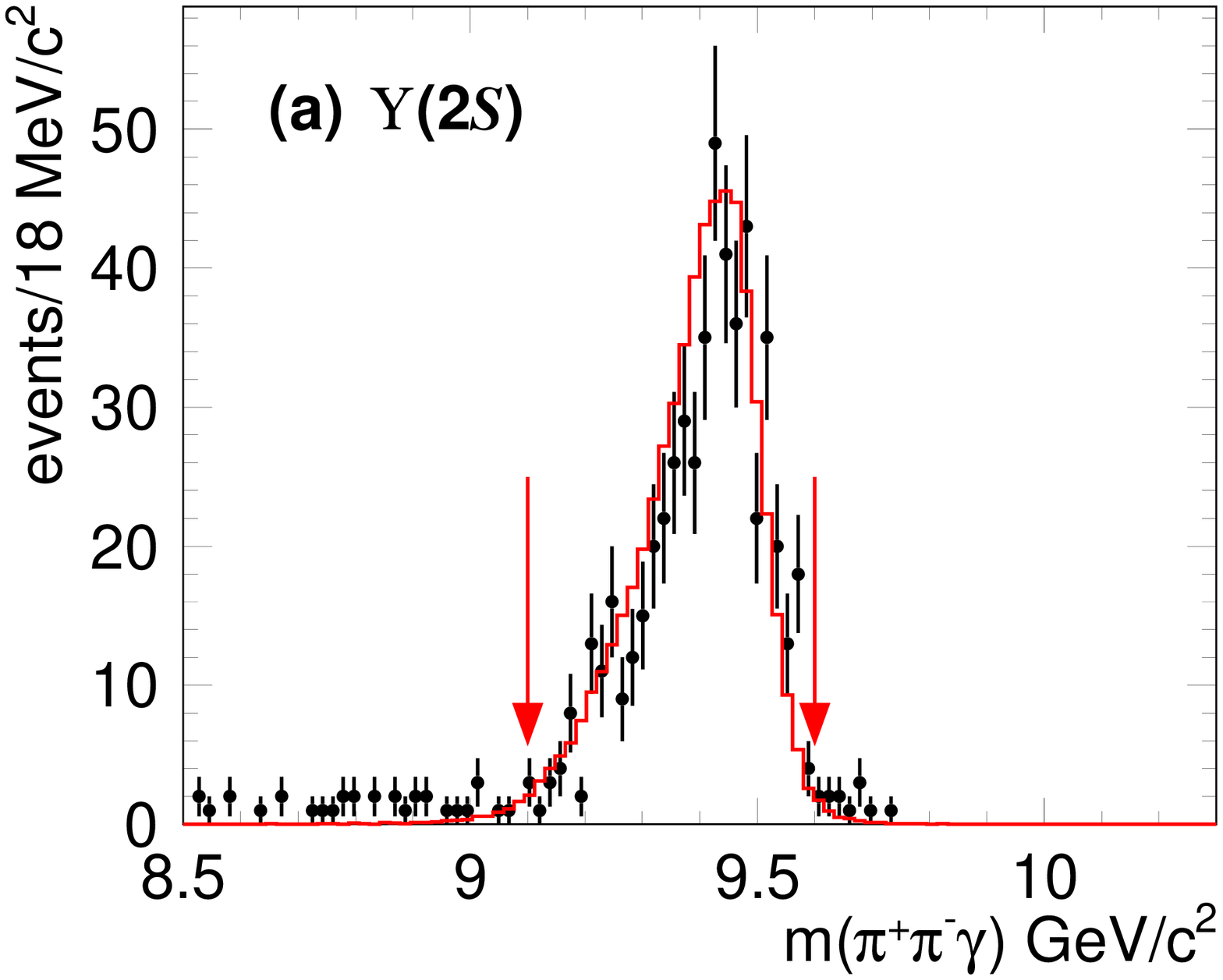}
\includegraphics[width=8.5cm]{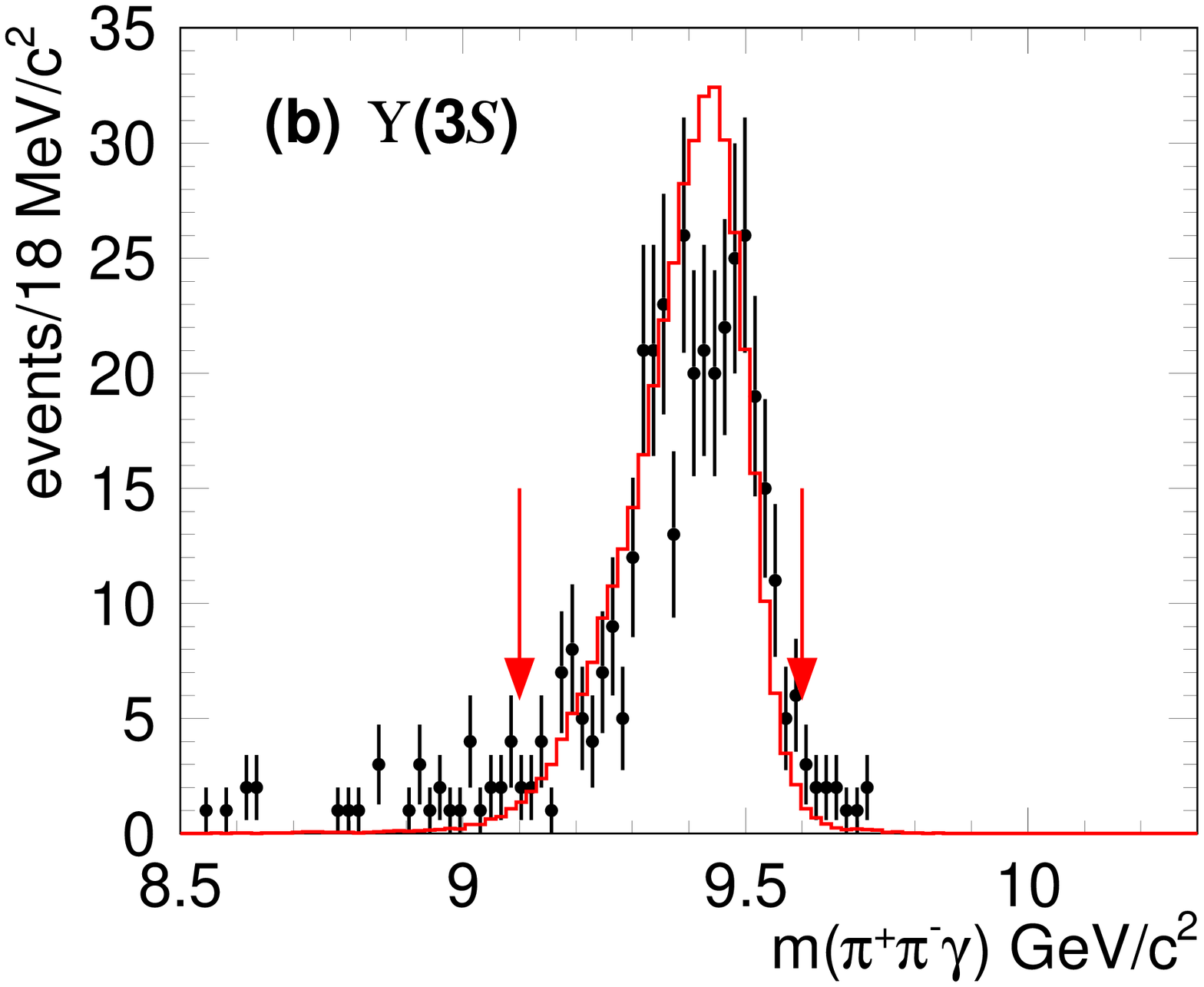}
\caption{$m(\gamma \pip \pim)$ mass distributions after the $\mm(\pips \pims)$
  selection for the
  (a) \TwoS and (b) \ThreeS data. The arrows indicate the range used to select the \OneS signal.
The full line histograms are the results from signal MC simulations.}
\label{fig:fig3}
\end{center}
\end{figure}

To reconstruct $\OneS \to \gamma \pip \pim$ decays,
we require a loose identification of both pions from the \OneS decay and
obtain the distributions of $m(\gamma \pip \pim)$ shown in
Fig.~\ref{fig:fig3}. The distributions show the expected peak at the \OneS mass
with little background but do not have a Gaussian shape due to the
asymmetric energy response of the EMC to a high-energy photon.
The full line histograms compare the data with signal MC simulations and show good agreement.

We finally isolate the decay $\OneS \to \gamma \pip \pim$  by requiring
\begin{equation}
9.1 \ \gevcc < m(\gamma \pip \pim)< 9.6 \ \gevcc.
\label{eq:mpipig}
\end{equation}
At this stage no more than one candidate per event is present.

\begin{figure}
\begin{center}
\includegraphics[width=8.5cm]{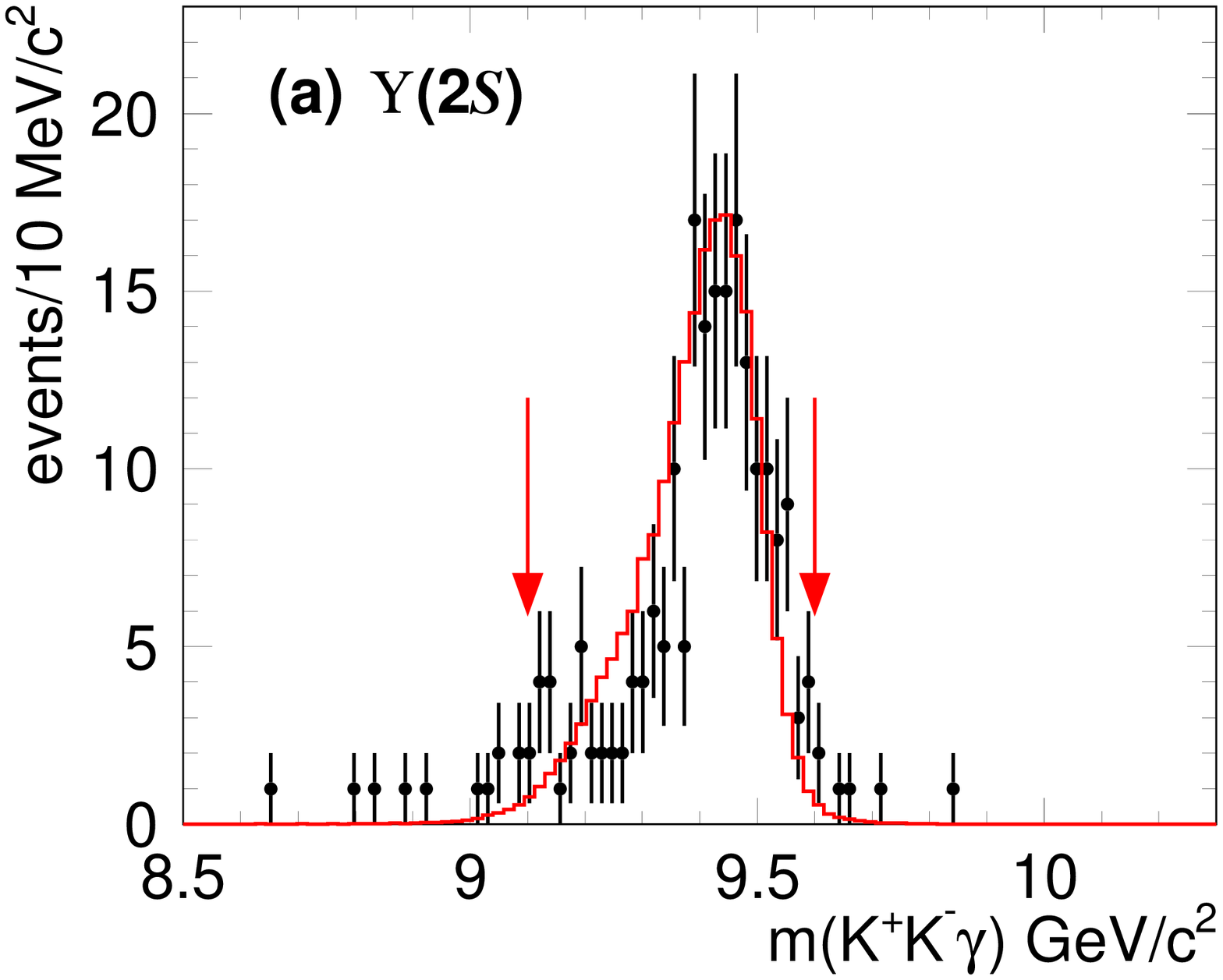}
\includegraphics[width=8.5cm]{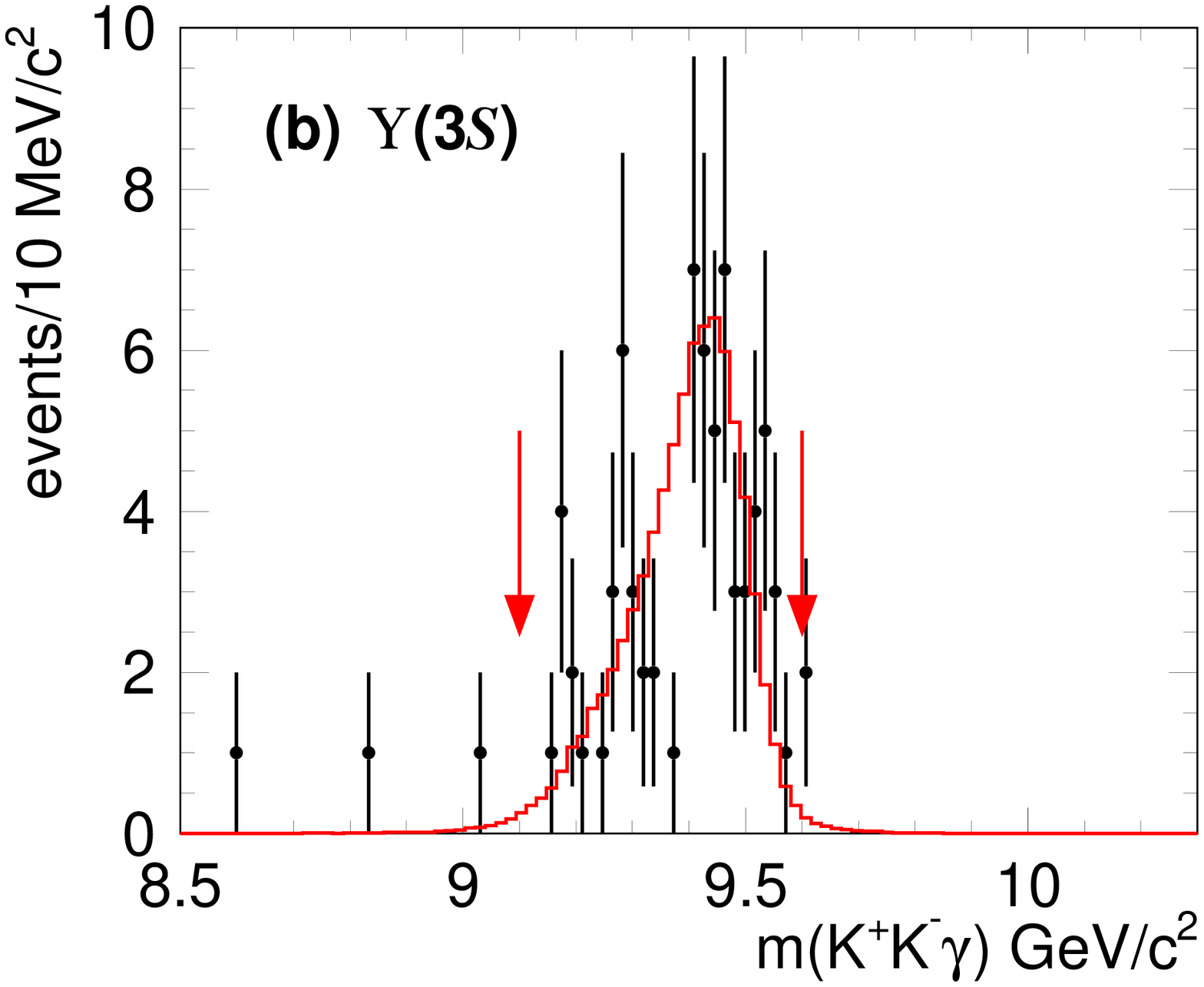}
\caption{$m(\kp \km \gamma)$ mass distributions after the $\mm(\pips \pims)$
  selection for the
 (a) \TwoS and (b) \ThreeS data. The arrows indicate the range used to select the \OneS signal.
The full line histograms are the results from signal MC simulations.}
\label{fig:fig4}
\end{center}
\end{figure}

We reconstruct the final state where $\OneS \to \gamma \kp
\km$ in a similar manner, by applying a loose identification of both kaons in the final state and
requiring the $m(\kp \km \gamma)$ mass, shown in Fig.~\ref{fig:fig4}, to be in the range

\begin{equation}
9.1 \ \gevcc < m(\kp \km \gamma)< 9.6 \ \gevcc.
\label{eq:mkkg}
\end{equation}

\section{\boldmath Study of the $\pip \pim$ and $\kp \km$ mass spectra}

\begin{figure}
\begin{center}
  \includegraphics[width=8.5cm]{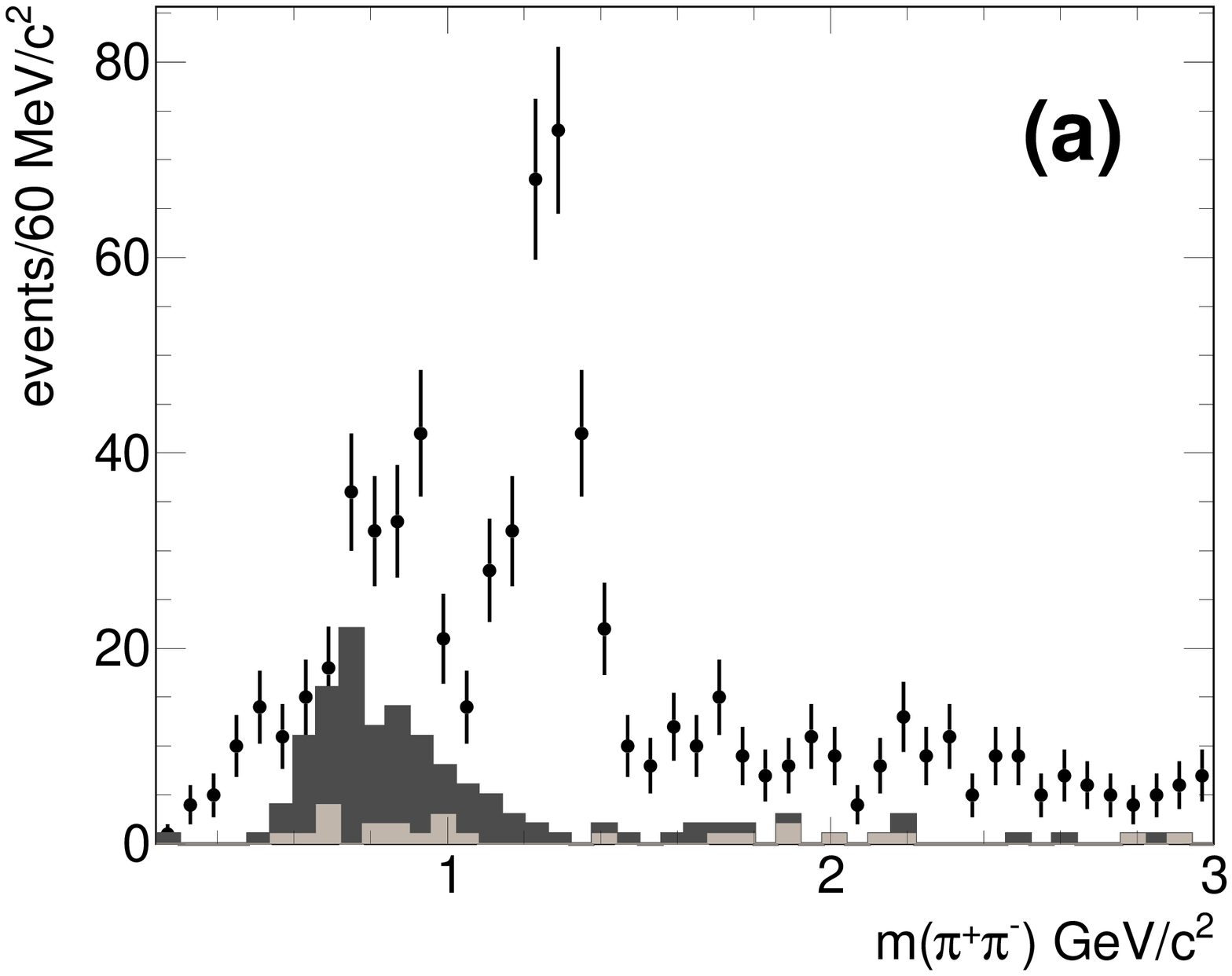}
  \includegraphics[width=8.5cm]{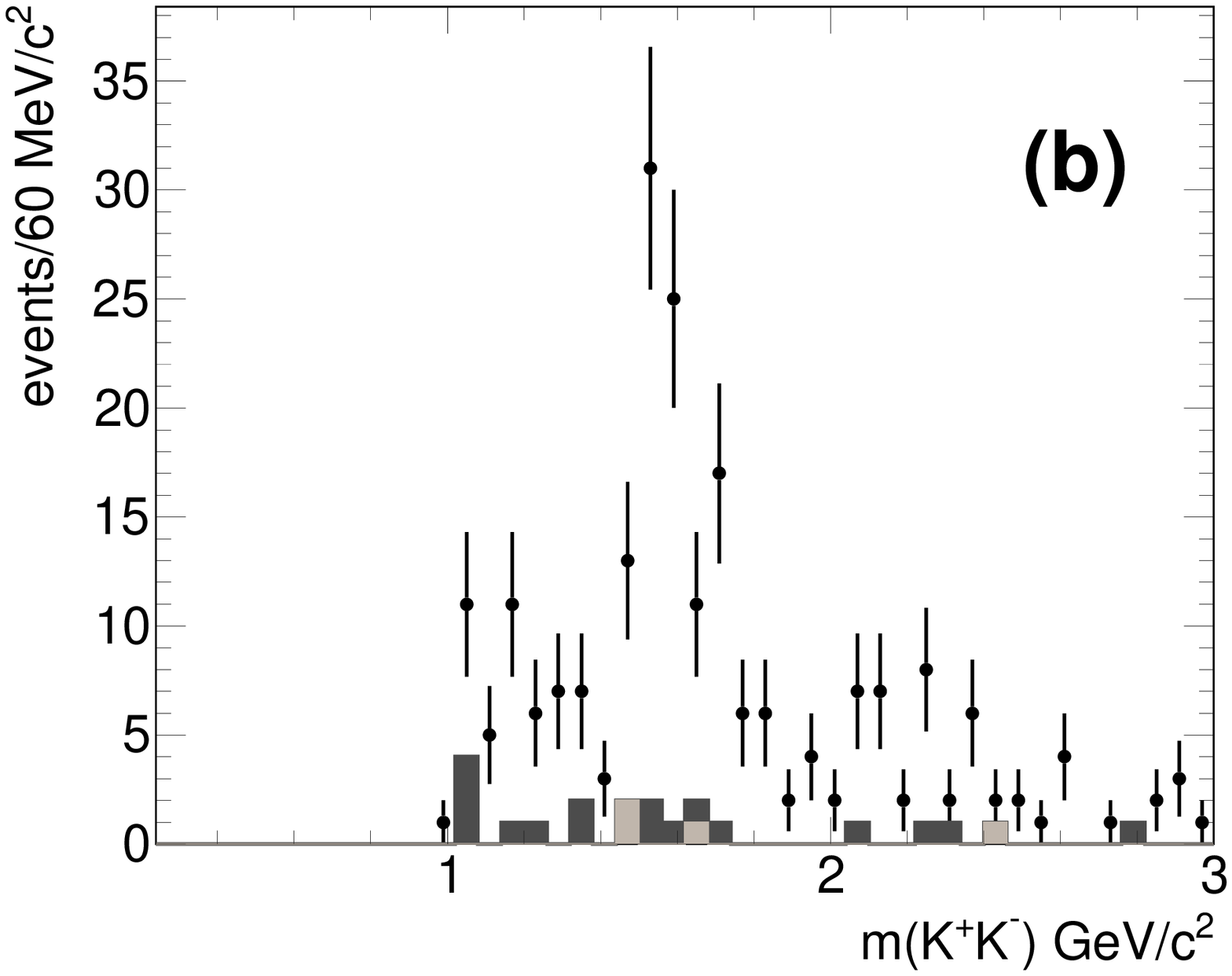}
\caption{(a) $\pip \pim$ mass distribution from $\OneS \to \pip \pim
  \gamma$ for the combined \TwoS and
  \ThreeS datasets. (b) $\kp \km$ mass distribution from $\OneS \to \kp \km
  \gamma$ for the combined \TwoS and
  \ThreeS datasets.
  The gray distributions show the expected background obtained from the corresponding $\mm(\pips \pims)$ sidebands.
The light-gray distributions evidences the background contribution from the \TwoS data.} 
\label{fig:fig5}
\end{center}
\end{figure}

The $\pip \pim$ mass spectrum, for $m(\pip \pim)<3.0$ \gevcc\
and summed over the \TwoS and \ThreeS datasets with 
507 and 277 events, respectively, is shown in Fig.~\ref{fig:fig5}(a).
The resulting $\kp \km$ mass spectrum, summed over the \TwoS and \ThreeS datasets with 
164 and 63 events, respectively, is shown in Fig.~\ref{fig:fig5}(b).
For a better comparison the two distributions are plotted using the same bin size and the same mass range.

We study the background for both $\pip \pim$ and $\kp \km$ final states using the $\mm(\pips \pims)$
sidebands. We select events in the $(4.5\sigma-7.0\sigma)$ regions on both sides of the signal region and require
the $m(\pip \pim \gamma)$ and $m(\kp \km \gamma)$ to be in the ranges defined by Eq.~\ref{eq:mpipig}
and Eq.~\ref{eq:mkkg}, respectively.
The resulting $\pip \pim$ and $\kp \km$ mass spectra for these events are superimposed in gray in Fig.~\ref{fig:fig5}(a) and
Fig.~\ref{fig:fig5}(b), respectively.
We note rather low background levels for all the final states, except for the $\pip \pim$ mass spectrum from the $\ThreeS$
data, which shows an enhancement at a mass of $\approx$ 750 \mevcc,
which we attribute to the presence of $\rho(770)^0$ background.
The $\pip \pim$ mass
spectrum from inclusive \ThreeS decays also shows a strong $\rho(770)^0$
contribution.

We search for background
originating from a possible hadronic $\OneS \to \pip \pim \piz$ decay, where
one of the two $\gamma$'s from the \piz\ decay is lost.
For this purpose, we make use of the \TwoS data and select events having four charged pions
and only one \piz\ candidate. We then select events satisfying
Eq.~(\ref{eq:mrec}) and plot the $\pip \pim \piz$ effective mass
distribution.
No \OneS signal is observed, which indicates that the branching
fraction for this possible \OneS decay mode is very small and
therefore that no contamination is expected 
in the study of the $\OneS \to \gamma \pip \pim$ decay mode.

\begin{figure}
\begin{center}
\includegraphics[width=8.5cm]{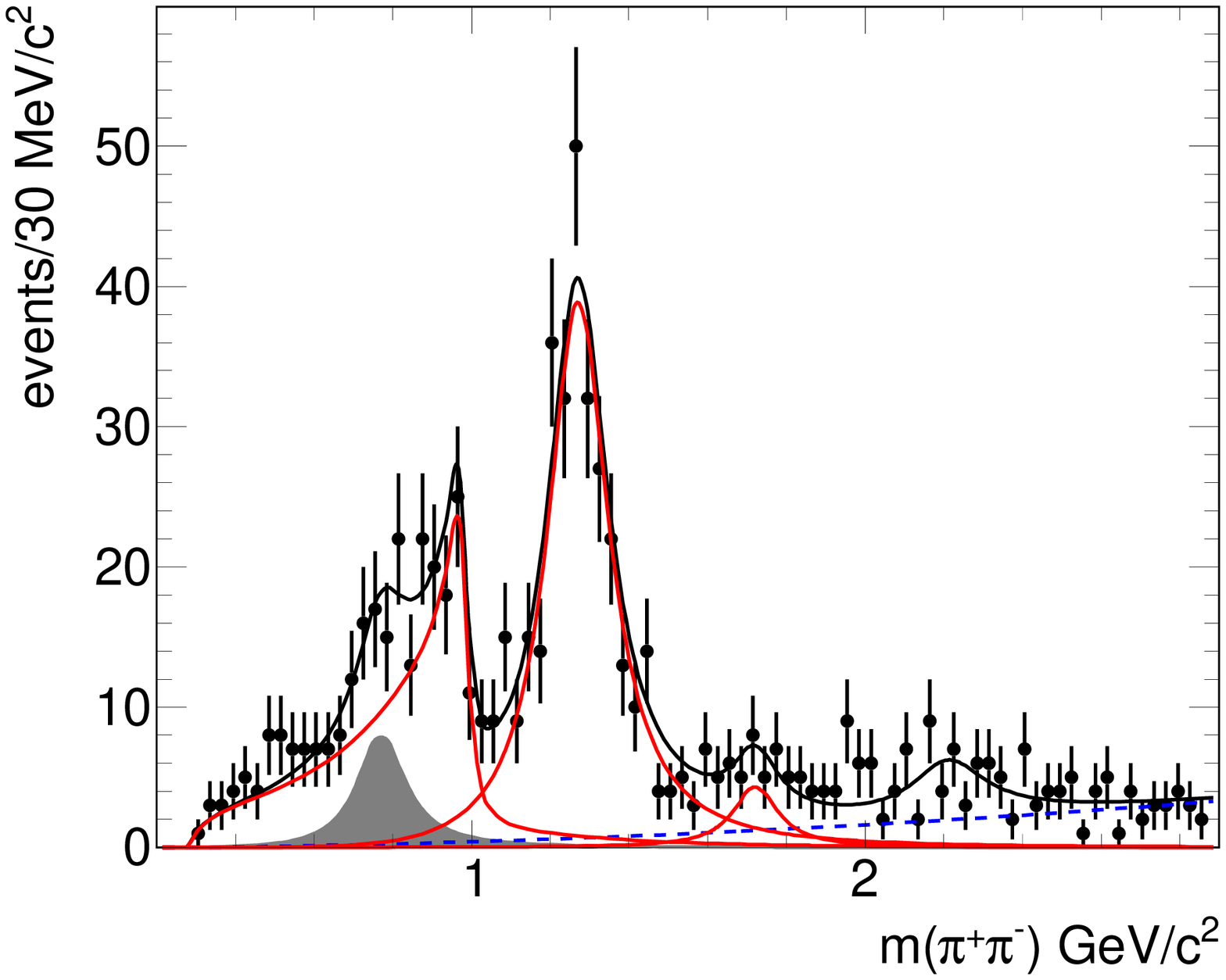}
\caption{$\pip \pim$ mass distribution from $\OneS \to \pip \pim
  \gamma$ for the combined \TwoS and
  \ThreeS datasets.
  The full (black) line is the results from the fit,
  the dashed (blue) line represents the fitted background.
The full (red) curves indicate the $S$-wave, $f_2(1270)$, and $f_0(1710)$ contributions.
  The shaded (gray) area
represents the fitted $\rho(770)^0$ background.}
\label{fig:fig6}
\end{center}
\end{figure}

\begin{figure}
\begin{center}
\includegraphics[width=8cm]{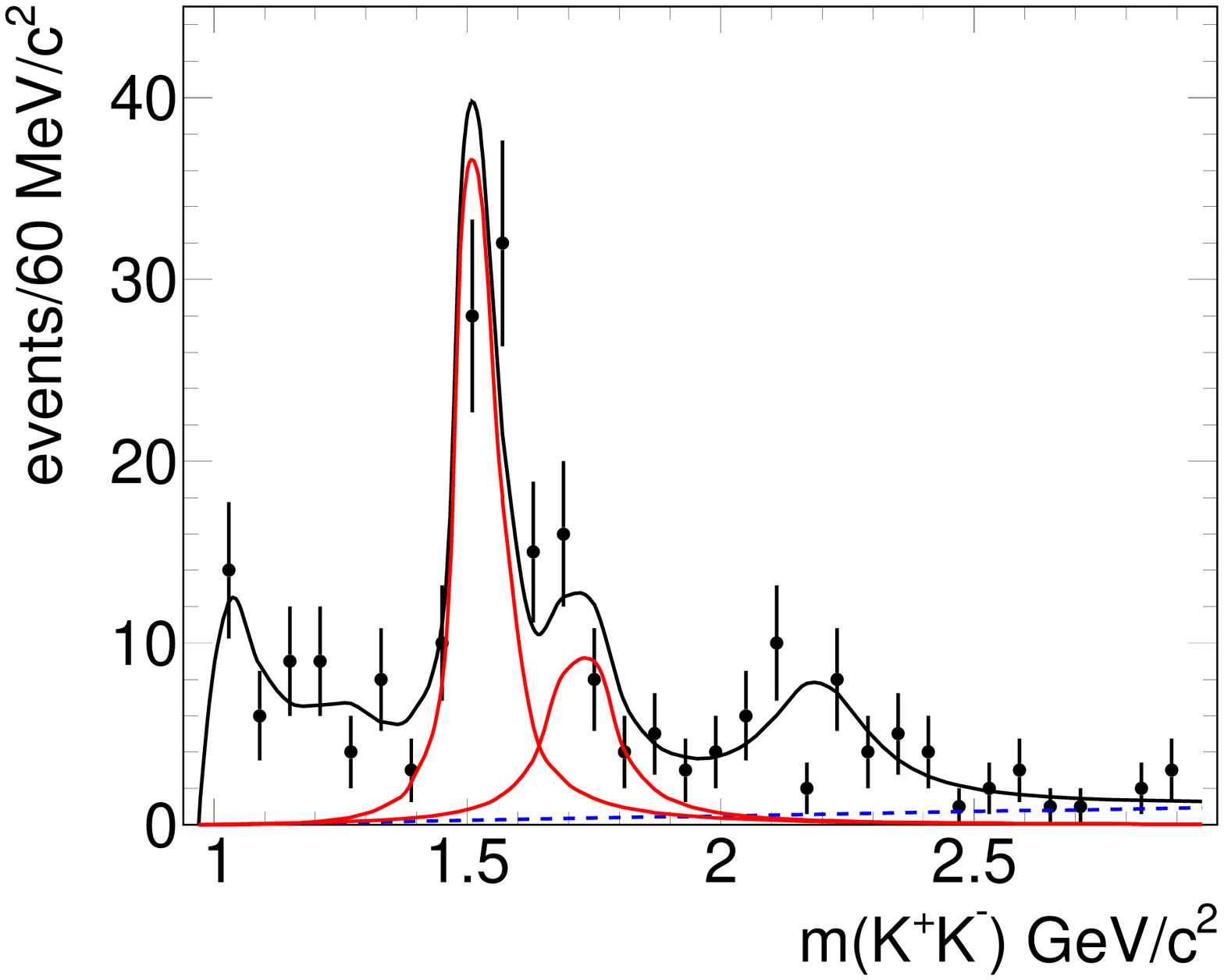}
\caption{$\kp \km$ mass distribution from $\OneS \to \kp \km
  \gamma$ for the combined \TwoS and
  \ThreeS datasets. The full (black) line is the results from the fit,
  the dashed (blue) line represents the fitted background.
The (red) curves show the contributions from $f_2'(1525)$ and $f_0(1710)$.}
\label{fig:fig7}
\end{center}
\end{figure}

The $\pip \pim$ mass spectrum, in 30 \mevcc\ bin size is shown in Fig.~\ref{fig:fig6}.
The spectrum shows $I=0$, $J^P={\rm even}^{++}$ resonance production, with low backgrounds above 1 \gevcc. We
observe a rapid drop around 1 \gevcc\ characteristic of the presence of the
$f_0(980)$, and a strong $f_2(1270)$ signal. The data also suggest the
presence of weaker resonant contributions.
The $\kp \km$ mass spectrum is shown in Fig.~\ref{fig:fig7} and also shows resonant production, with
low background.
Signals at the positions of $f_2'(1525)$ and $f_0(1710)$ can be
observed.

We make use of a phenomenological model to extract the different $\OneS \to \gamma R$ branching fractions,
where $R$ is an intermediate resonance.

\subsection{\boldmath Fit to the $\pip \pim$ mass spectrum}

We perform a simultaneous binned fit to the $\pip \pim$ mass spectra from the \TwoS and \ThreeS datasets  using the following model.
\begin{table*}[htb]
  \caption{Resonances yields and statistical significances from the
    fits to the $\pi^+ \pi^-$ and $\Kp \Km$ mass spectra for the \TwoS
    and \ThreeS datasets. The symbol $f_J(1500)$ indicates the signal in the 1500 \mevcc\
    mass region.
When two errors are reported, the first is statistical and the second
systematic.
Systematic uncertainties are evaluated only for resonances for which
we compute branching fractions.}
   \label{tab:br1}
\begin{center}
\begin{tabular}{lccc}
  \hline
\noalign{\vskip2pt}
Resonances ($\pip \pim$)&  Yield \TwoS &  Yield \ThreeS & Significance\cr
\hline
\noalign{\vskip2pt}
$S$-wave & $ 133 \pm 16 \pm 13$ &  $87 \pm 13$  & 12.8$\sigma$ \cr
$f_2(1270)$ & $255 \pm 19 \pm \al 8$ & $77 \pm 7 \pm 4$ & 15.9$\sigma$\cr
$f_0(1710)$ & $ \al 24 \pm \al 8 \pm \al 6$ & $\al 6  \pm 8 \pm 3$ & \al\all 2.5$\sigma$\cr 
$f_0(2100)$ & $33 \pm 9$ & $ \al 8 \pm 15$ & \cr 
$\rho(770)^0$ &         & 54 $\pm$ 23 & \cr
\hline
\noalign{\vskip2pt}
Resonances ($\Kp \Km$)&  Yield \TwoS + \ThreeS &  & Significance \cr
\hline
\noalign{\vskip2pt}
$f_0(980)$ & $47 \pm \al 9 $ &   & 5.6$\sigma$ \cr 
$f_J(1500)$ & $77 \pm 10 \pm 10$ &  & 8.9$\sigma$ \cr
$f_0(1710)$ & $ 36 \pm 9 \pm \al 6 $  & & 4.7$\sigma$ \cr 
$f_2(1270)$ & $15 \pm 8$ &  &  \cr
$f_0(2200)$ & $38 \pm 8$ & \cr
\hline
\end{tabular}
\end{center}
\end{table*}
\begin{itemize}
\item{}
  We describe the low-mass region (around the $f_0(500)$)
  using a relativistic $S$-wave Breit-Wigner lineshape having free parameters.
  We test the $S$-wave hypothesis in Sec.VI and Sec.VIII.
 We obtain its parameters from the \TwoS data only, and we fix them 
in the description of the \ThreeS data.
\item{}
We describe the $f_0(980)$ using the Flatt\'{e}~\cite{flatte} formalism.
For the $\pi^+\pi^-$ channel the Breit-Wigner lineshape has the
form
\begin{equation}
BW(m) = \frac{m_0 \sqrt{\Gamma_i} \sqrt{\Gamma_\pi(m)}}
{m_0^2 - m^2 -im_0(\Gamma_\pi(m) + \Gamma_K(m))},
\label{eq:fla_pipi}
\end{equation}
and in the $K^+ K^-$ channel the Breit-Wigner function has the form
\begin{equation}
BW(m) = \frac{m_0 \sqrt{\Gamma_i} \sqrt{\Gamma_K(m)}}
{m_0^2 - m^2 -im_0(\Gamma_\pi(m) + \Gamma_K(m))},
\label{eq:fla_kk}
\end{equation}
where $\Gamma_i$ is absorbed into the intensity of the resonance.
$\Gamma_\pi(m)$ and
$\Gamma_K(m)$ describe the partial widths of the resonance to decay to $\pi \overline \pi$ and $K \overline K$
and are given by
\begin{equation}
\begin{array}{l}
\Gamma_\pi(m) = g_\pi(\frac{m^2}{4} - m^2_\pi)^{1/2}, \\
\Gamma_K(m) = \frac{g_K}{2}[(\frac{m^2}{4} - m^2_{K^+})^{1/2} + (\frac{m^2}{4}-m^2_{K^0})^{1/2} ],
\end{array}
\end{equation}
where $g_\pi$ and $g_K$ are the squares of the coupling constants of the
resonance
to the $\pi \overline \pi $ and $K \overline K$ systems. The
$f_0(980)$ parameters and couplings are taken from
Ref.~\cite{Armstrong}: $m_0=0.979\pm 0.004$ \gevcc, $g_\pi=0.28 \pm 0.04$ and $g_K=0.56 \pm 0.18$.
  \item{}
    The total $S$-wave is described by a coherent sum of $f_0(500)$ and $f_0(980)$ as
    \begin{equation}
\begin{array}{l}
      S\text{-wave} = \\
    \mid BW_{f_0(500)}(m) + c BW_{f_0(980)}(m)e^{i \phi} \mid^2.
\end{array}
    \end{equation}
    where $c$ and $\phi$ are free parameters for the relative intensity and phase of the two interfering contributions.
  \item{} The $f_2(1270)$ and $f_0(1710)$ resonances are represented
    by relativistic Breit-Wigner functions with parameters fixed to
    PDG values~\cite{PDG}.
  \item{} In the high $\pip \pim$ mass region
we are unable, with the present statistics, to distinguish the
different possible resonant contributions. 
Therefore we make use of the method used by CLEO~\cite{Dobbs} and include a single resonance $f_0(2100)$ having a width fixed to the PDG value
($224 \pm 22$) and unconstrained mass.
  \item{}
    The background is parametrized with a quadratic dependence
    $$b(m)=p(m)(a_1m + a_2m^2),$$
where $p(m)$ is the $\pi$ center-of-mass momentum in the
$\pip \pim$ rest frame, which goes to zero at $\pip \pim$ threshold.
\item{} For the \ThreeS data we also include $\rho(770)^0$ background with parameters fixed to the PDG values.
\end{itemize}
The fit is shown in Fig.~\ref{fig:fig6}. It has 16 free parameters and $\chi^2=182$ for ndf=152, corresponding to a $p$-value of 5\%.
The yields and statistical significances are reported in Table~\ref{tab:br1}.
Significances are computed as follows: for each resonant contribution (with fixed parameters) we set the yield to zero
and compute the significance as $\sigma=\sqrt{\Delta \chi^2}$, where $\Delta \chi^2$ is the difference 
in $\chi^2$ between the fit with and without the presence of the resonance.

The table also reports systematic uncertainties on the yields,
evaluated as follows: the parameters of each resonance are modified according to $\pm \sigma$, where $\sigma$ is the PDG uncertainty and
the deviations from the reference fit are added in quadrature.
The background has been modified to have a linear shape.
The effective range in the Blatt-Weisskopf~\cite{blatt} factors entering in the description of the intensity and the width of
the relativistic Breit-Wigner function have been varied between 1 and 5 $\gev^{-1}$, and
the average deviation is taken as a systematic uncertainty.
The different contributions, dominated by the uncertainties on the resonances parameters, are added in quadrature.

We note the observation of a significant $S$-wave in \OneS radiative decays. This observation
was not possible in the study of $\jpsi$ radiative decay to $\pip \pim$ because of the presence of a strong,
irreducible background from $\jpsi \to \pip \pim \piz$~\cite{pipi_markiii}.
We obtain the following $f_0(500)$ parameters:
\begin{equation}
\begin{array}{l}
  m(f_0(500)) = 0.856 \pm 0.086 \ \gevcc, \\
\Gamma(f_0(500))=1.279 \pm 0.324 \ \gev, 
\end{array}
\end{equation}
and $\phi=2.41 \pm 0.43$ rad. The fraction of $S$-wave events associated with the $f_0(500)$ is $(27.7\pm3.1)\%$.
We also obtain $m(f_0(2100))= 2.208 \pm 0.068 \ \gevcc$.

\subsection{Study of the $\Kp \Km$ mass spectrum.}

Due to the limited statistics we do not separate the data into the \TwoS and \ThreeS datasets.
We perform a binned fit to the combined $\Kp \Km$ mass spectrum using the following model:

\begin{itemize}
  \item{}
    The background is parametrized with a linear dependence starting with zero at threshold.
    \item{}
    The $f_0(980)$ is parametrized according to the  Flatt\'{e} formalism
    described by Eq.~(\ref{eq:fla_kk}) for the $\kp \km$ projection.
  \item{} The $f_2(1270)$, $f_2'(1525)$, $f_0(1500)$, and $f_0(1710)$ resonances are
    represented by relativistic Breit-Wigner functions with parameters
    fixed to PDG values. 
\item{} We include an $f_0(2200)$ contribution having parameters fixed to the PDG values.
\end{itemize}

The fit shown in Fig.~\ref{fig:fig7}. It has six free parameters and $\chi^2=35$ for ndf=29,
corresponding to a $p$-value of 20\%; the yields and significances are reported in Table~\ref{tab:br1}.
Systematic uncertainties have been evaluated as for the fit to the $\pip \pim$ mass spectrum.
The parameters of each resonance are modified according to $\pm \sigma$, where $\sigma$ is the PDG uncertainty and
the deviations from the reference fit are added in quadrature.
The background has been modified to have a quadratic shape.
The effective range in the Blatt-Weisskopf~\cite{blatt} factors entering in the description of the intensity and the width of
the relativistic Breit-Wigner function have been varied between 1 and 5 $\gev^{-1}$, and
the average deviation is taken as a systematic uncertainty.
The different contributions, dominated by the uncertainties on the resonances parameters, are added in quadrature.
In the 1500 \mevcc\ mass region both $f_2'(1525)$ and $f_0(1500)$ can contribute, therefore
we first fit the mass spectrum assuming the presence of $f_2'(1525)$ only and then replace in the fit the $f'_2(1525)$ with the $f_0(1500)$ resonance.
In Table~\ref{tab:br1} we label this contribution as $f_J(1500)$.
The resulting yield variation between the two fits is small and gives a negligible contribution to the total systematic uncertainty.
A separation of the $f'_2(1525)$ and $f_0(1500)$ contributions is discussed in Secs. VI and VII.

\section{\boldmath Efficiency correction}

\subsection{Reconstruction efficiency}

\begin{figure}[!htb]
\begin{center}
  \includegraphics[width=9cm]{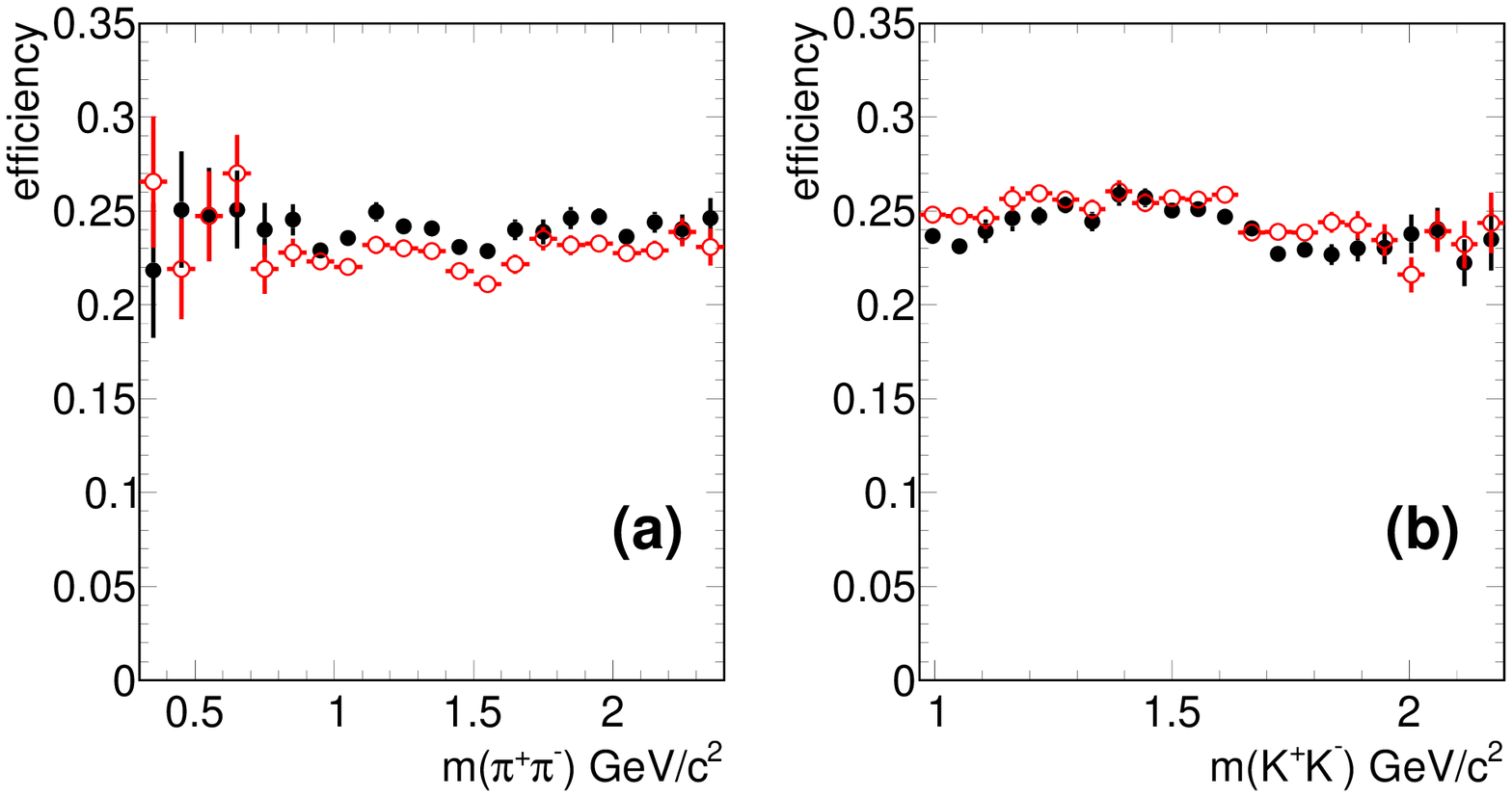}
  \caption{Efficiency distributions as function of mass for the \TwoS/\ThreeS data. (a) $\pip \pim \gamma$, (b) $\kp \km \gamma$.
    Filled (black) circles are for \TwoS data, open (red) circles are for \ThreeS data.}
\label{fig:fig8}
\end{center}
\end{figure}

To compute the efficiency, MC signal events are generated using a
detailed detector simulation~\cite{geant}.
These simulated events are reconstructed and analyzed in the same manner as data. The efficiency is computed as the ratio between 
reconstructed and generated events.
The efficiency distributions as functions of mass, for the \TwoS/\ThreeS data and for the $\pip \pim \gamma$ and $\kp \km \gamma$ final states, are
shown in fig.~\ref{fig:fig8}. We observe an almost uniform behavior for all the final states.

We define the helicity angle $\theta_H$ as the angle formed by the $h^+$ (where $h=\pi,K$), in the $h^+ h^-$ rest frame, and the
$\gamma$ in the $h^+ h^- \gamma$ rest frame.
We also define $\theta_{\gamma}$ as the angle formed by the radiative photon in the $h^+ h^- \gamma$ rest frame with respect to the
\OneS direction in the \TwoS/\ThreeS rest frame.

We compute the efficiency in two different ways.
\begin{itemize}
\item{}
We label with $\epsilon(m,\cos \theta_H)$ the efficiency  computed as a
function of the $h^+ h^-$ effective mass and the helicity angle $\cos \theta_H$. 
This is used only to obtain efficiency-corrected mass spectra.
\item{}
We label with $\epsilon(\cos \theta_H,\cos \theta_{\gamma})$ the efficiency computed, for each resonance mass window (defined
in Table ~\ref{tab:pwa_h}), as a function of
$\cos \theta_H$ and $\cos \theta_{\gamma}$. This is used to obtain the
efficiency-corrected 
angular distributions and branching fractions of the different
resonances. 
\end{itemize}
 
To smoothen statistical fluctuations in the evaluation of $\epsilon(m,\cos \theta_H)$, for $\OneS \to \gamma
\pip \pim$, we divide the $\pip \pim$ mass into nine 300-\mevcc-wide
intervals and plot the $\cos \theta_H$ in
each interval. The distributions of $\cos \theta_H$ are then fitted
using cubic splines~\cite{spline}. The efficiency at each $m(\pip
\pim)$ is then computed using a linear
interpolation between adjacent bins.

\begin{figure}[!htb]
\begin{center}
  \includegraphics[width=9cm]{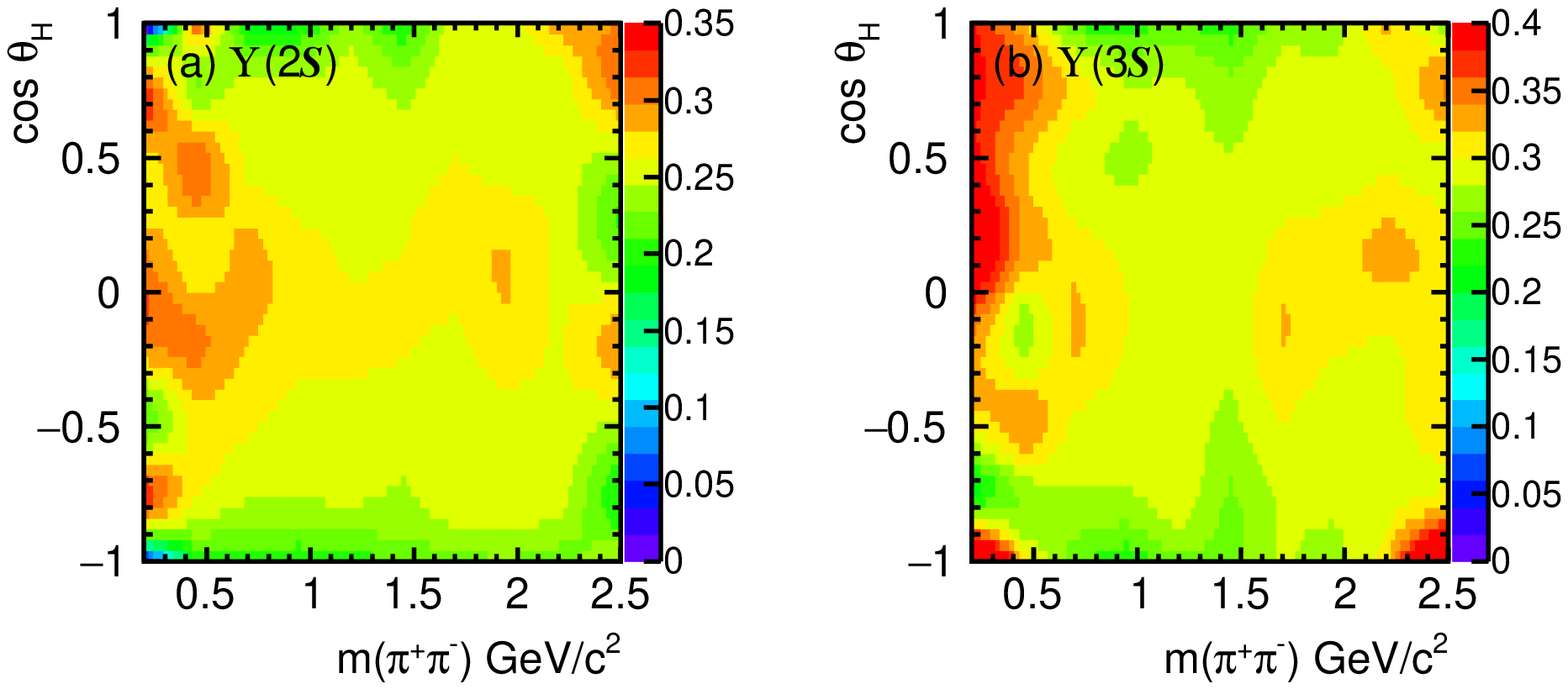}
 \caption{Fitted efficiency distribution in the ($m(\pip \pim)$, $\cos
   \theta_H$) plane for $\OneS \to \gamma \pip \pim$ for the (a) \TwoS
   and (b) \ThreeS datasets.}
\label{fig:fig9}
\end{center}
\end{figure}

Figure~\ref{fig:fig9} shows the efficiency distributions in the ($m(\pip\pim)$, $\cos \theta_H$) plane for the \TwoS and \ThreeS datasets.
We observe an almost uniform behavior with some loss at $\cos\theta_H$
close to $\pm 1$.
The efficiencies integrated over $\cos \theta_H$ are consistent with
being constant
with mass and have average values of $\epsilon(\TwoS \to \pip \pim \OneS (\to \gamma \pip \pim) )= 0.237 \pm 0.001$
and $\epsilon(\ThreeS \to \pip \pim \OneS (\to \gamma \pip \pim)) = 0.261 \pm 0.001$.

\begin{figure}[!htb]
\begin{center}
  \includegraphics[width=9cm]{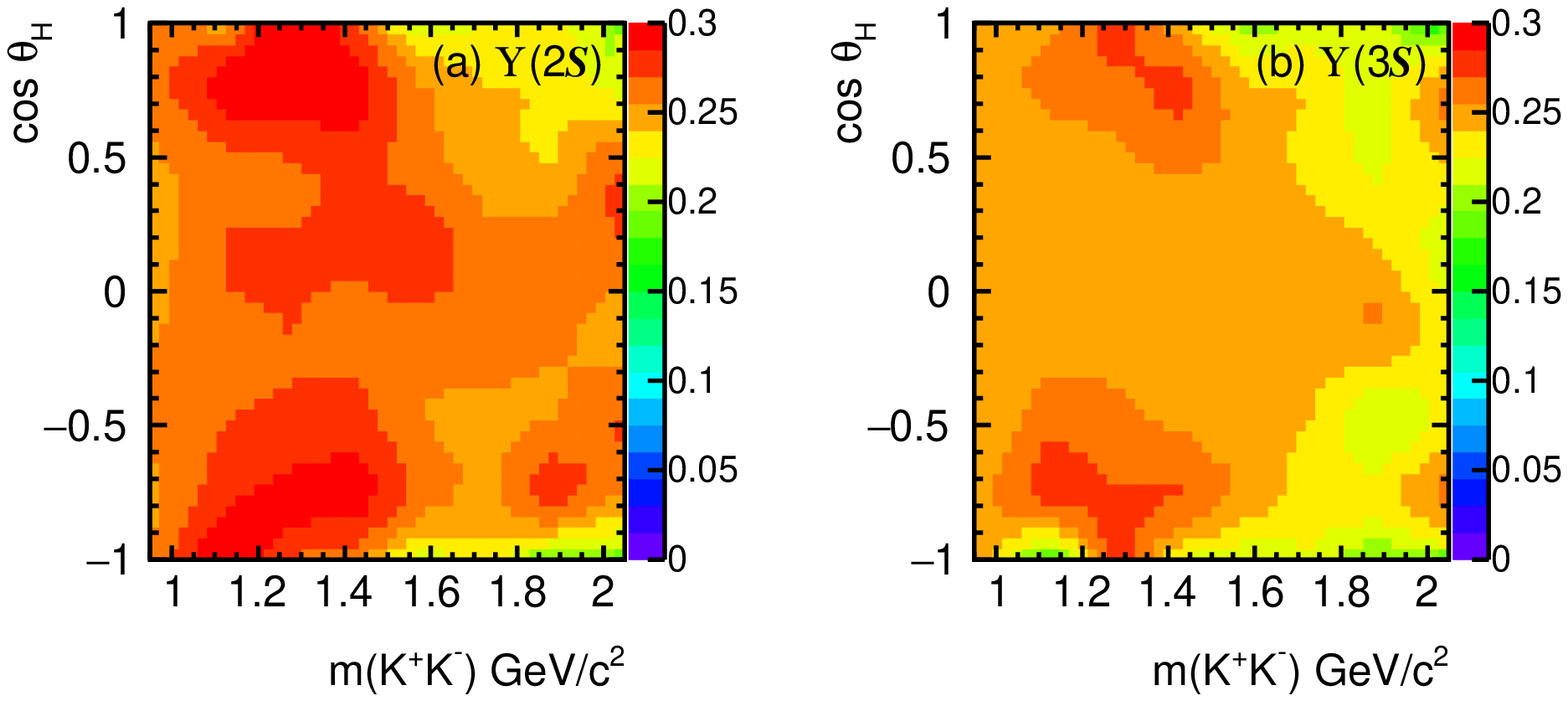}
 \caption{Fitted efficiency distribution in the ($m(\kp \km)$, $\cos
   \theta_H$) plane for  $\OneS \to \gamma \kp \km$ for the (a) \TwoS
   and (b) \ThreeS datasets.}
\label{fig:fig10}
\end{center}
\end{figure}

A similar method is used to compute $\epsilon(m,\cos \theta_H)$
for the $\OneS \to \gamma \kp \km$ final state. The average efficiency values are 
$\epsilon(\TwoS \to \pip \pim \OneS (\to \Kp \Km \gamma) )= 0.241 \pm
0.001$ and  $\epsilon(\ThreeS \to \pip \pim \OneS (\to \Kp \Km \gamma))
= 0.248 \pm 0.001$.
Figure~\ref{fig:fig10} shows the efficiency distributions in the ($m(\kp
\km)$, $\cos \theta_H$) plane for the \TwoS and \ThreeS datasets.

\begin{figure}[!htb]
\begin{center}
\includegraphics[width=8.5cm]{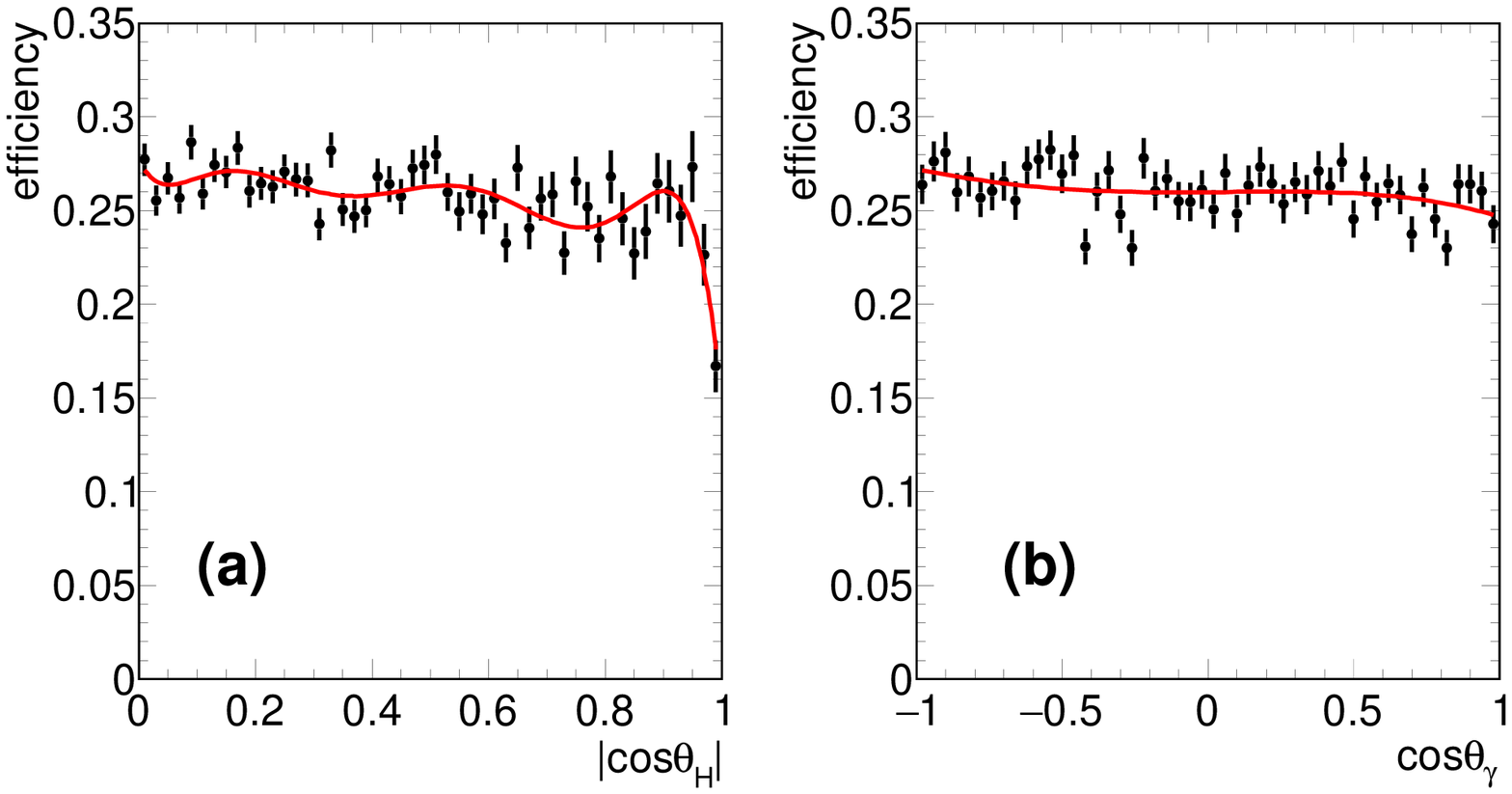}
\includegraphics[width=8.5cm]{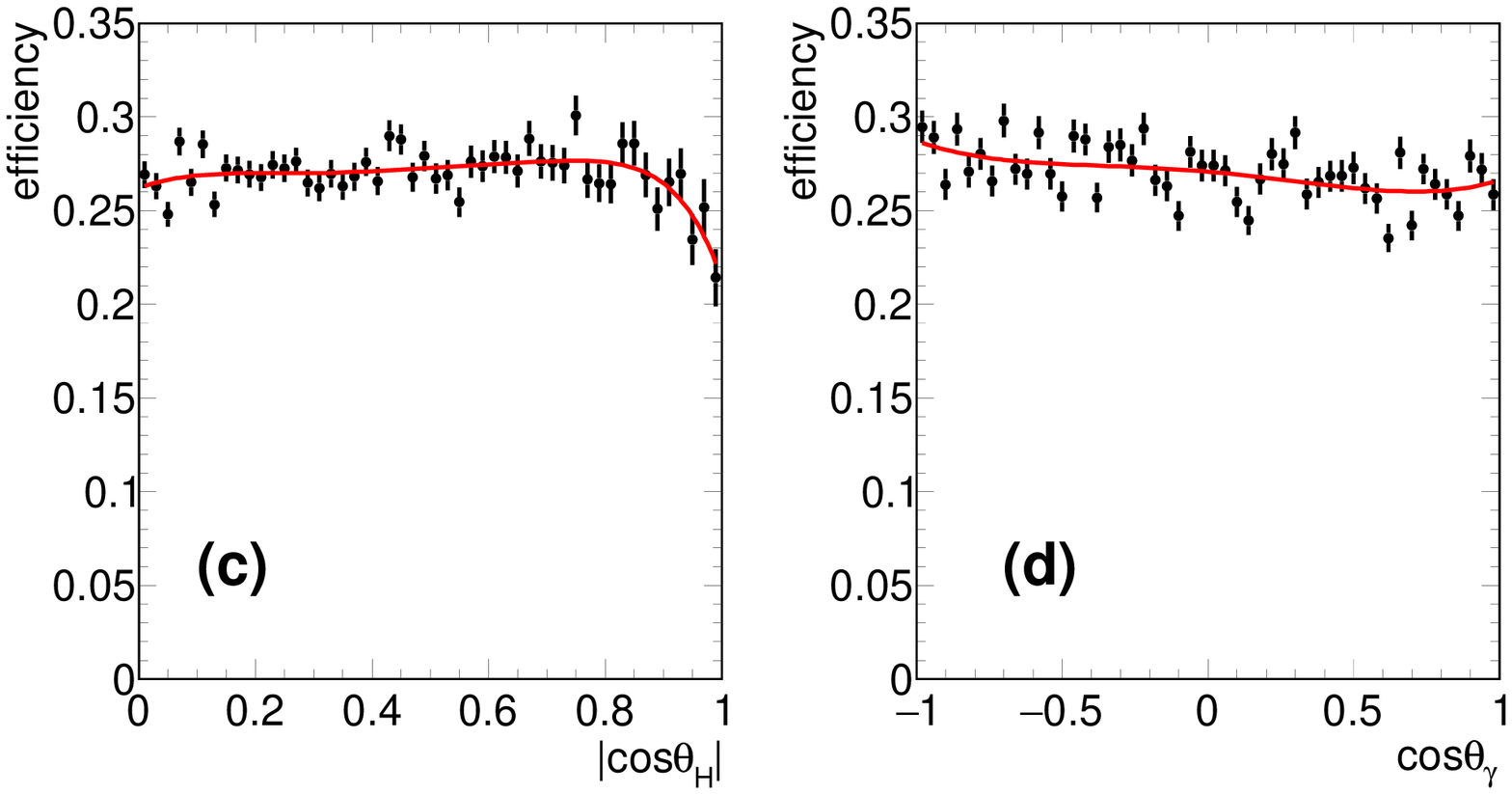}
\caption{Efficiency as a function of (a) $\cos \theta_H$ and (b)
  $\cos \theta_{\gamma}$ for
$\TwoS \to \pi^+_s \pi^-_s \OneS \to \gamma f_2(1270)(\to \pip \pim)$. 
Efficiency as a function of (c) $\cos \theta_H$ and (d)
  $\cos \theta_{\gamma}$ for $\TwoS \to \pi^+_s \pi^-_s \OneS \to \gamma f_2'(1525)(\to \kp \km)$. 
The lines are the result of the polynomial fits.
} 
\label{fig:fig11}
\end{center}
\end{figure}

We also compute the efficiency in the $(\cos \theta_H,\cos \theta_{\gamma})$ plane
for each considered resonance decaying to $\pip \pim$ and $\kp \km$.
Since there are no correlations between these two variables,
we parametrize the efficiency as
\begin{equation}
\epsilon(\cos \theta_H,\cos \theta_{\gamma})=\epsilon(\cos \theta_H)\times\epsilon(\cos \theta_{\gamma}).
\label{eq:eff}
\end{equation}
The distributions of the efficiencies as functions of $\cos \theta_H$
and $\cos \theta_{\gamma}$ are shown in Fig.~\ref{fig:fig11} for the
$f_2(1270) \to \pip \pim$ and $f_2'(1525) \to \kp \km$ mass regions,
for the \TwoS datasets. To smoothen statistical fluctuations, the
efficiency projections are fitted using 7-th and 4-th order polynomials, respectively.
Similar behavior is observed for the other resonances and for the
\ThreeS datasets.
\begin{table*}[htb]
  \caption{Efficiency corrections and efficiency corrected yields for each resonance and dataset.
    The symbol $f_J(1500)$ indicates the signal in the 1500 \mevcc\
    mass region.
    The error on the efficiency weight $w_R$ includes all the systematic uncertainties related to the reconstruction.
  The events yields are presented with statistical and total systematic uncertainties.}
    \label{tab:br2}
\begin{center}
\begin{tabular}{lcccc}
  \hline
\noalign{\vskip2pt}
Resonance & \TwoS &\TwoS & \ThreeS & \ThreeS \cr
                                                             
$\pip \pim$ & $w_R$ & corrected yield & $w_R$ & corrected
                                                             yield \cr
\hline
\noalign{\vskip2pt}
 $S$-wave     & $4.07 \pm 0.06$ & $\al 541 \pm 65 \pm 53$ &  \cr
$f_2(1270)$ & $4.09 \pm 0.06$  & $1043 \pm 78 \pm 36$ & $3.70 \pm 0.05$
& $285 \pm 26 \pm 15$ \cr
$f_0(1710)$ & $3.97 \pm 0.17$ & $\al\al 95 \pm 32 \pm 24$ & $3.60 \pm 0.08$ &
$\al 22 \pm 29 \pm 11$ \cr
\hline
\noalign{\vskip2pt} 
Resonance & \TwoS/\ThreeS & \TwoS/\ThreeS  & & \cr
$\kp \km$                  & $w_R$ & corrected yield &  &  \cr
\hline
\noalign{\vskip2pt}
$f_J(1500)$ & $3.65 \pm 0.14$ & $281 \pm 37 \pm 38$  & & \cr
$f_0(1710)$ & $3.96 \pm 0.13$ & $143 \pm 36 \pm 24$ &  & \cr
\hline
\end{tabular}
\end{center}
\end{table*}

\subsection{Efficiency correction}

To obtain the efficiency correction weight $w_R$ for the resonance $R$ we
divide each event by the efficiency $\epsilon(\cos \theta_H,\cos \theta_{\gamma})$

\begin{equation}
w_R = \frac{\sum_{i=1} ^{N_R}1/\epsilon_i(\cos \theta_H,\cos \theta_{\gamma})}{N_R},
\label{eq:weff}
\end{equation}
\noindent where $N_R$ is the number of events in the resonance mass range. 
The resulting efficiency weight for each resonance is reported in
Table~\ref{tab:br2}. We compute separately the \TwoS and \ThreeS yields 
for resonances decaying to $\pip \pim$ while, due to the limited
statistics,
for resonances decaying to $\kp \km$ the two datasets are merged and
corrected using the weighted average efficiency. 
The systematic effect related to the effect of particle
identification is assessed by the use of high statistics control
samples. We assign systematic uncertainties of 0.2\% to the identification of each pion and 1.0\% to that of each kaon.
We include an efficiency correction of $0.9885 \pm 0.0065$
to the reconstruction of the high energy photon, obtained from studies on Data/MC detection efficiency.
The efficiency correction contribution due to the limited MC statistics
is included using the statistical
uncertainty on the average efficiency weight as well as the effect of the fitting procedure.
The above effects are added in quadrature and are presented in Table~\ref{tab:br2} as systematic uncertainties related
to the efficiency correction weight.
Finally we propagate
the systematic effect on event yields obtained from the fits to the
mass spectra. The resulting efficiency corrected yields are reported in Table~\ref{tab:br2}.

\section{Legendre Polynomial Moments analysis}

\begin{figure*}[!htb]
\begin{center}
\includegraphics[width=8.0cm]{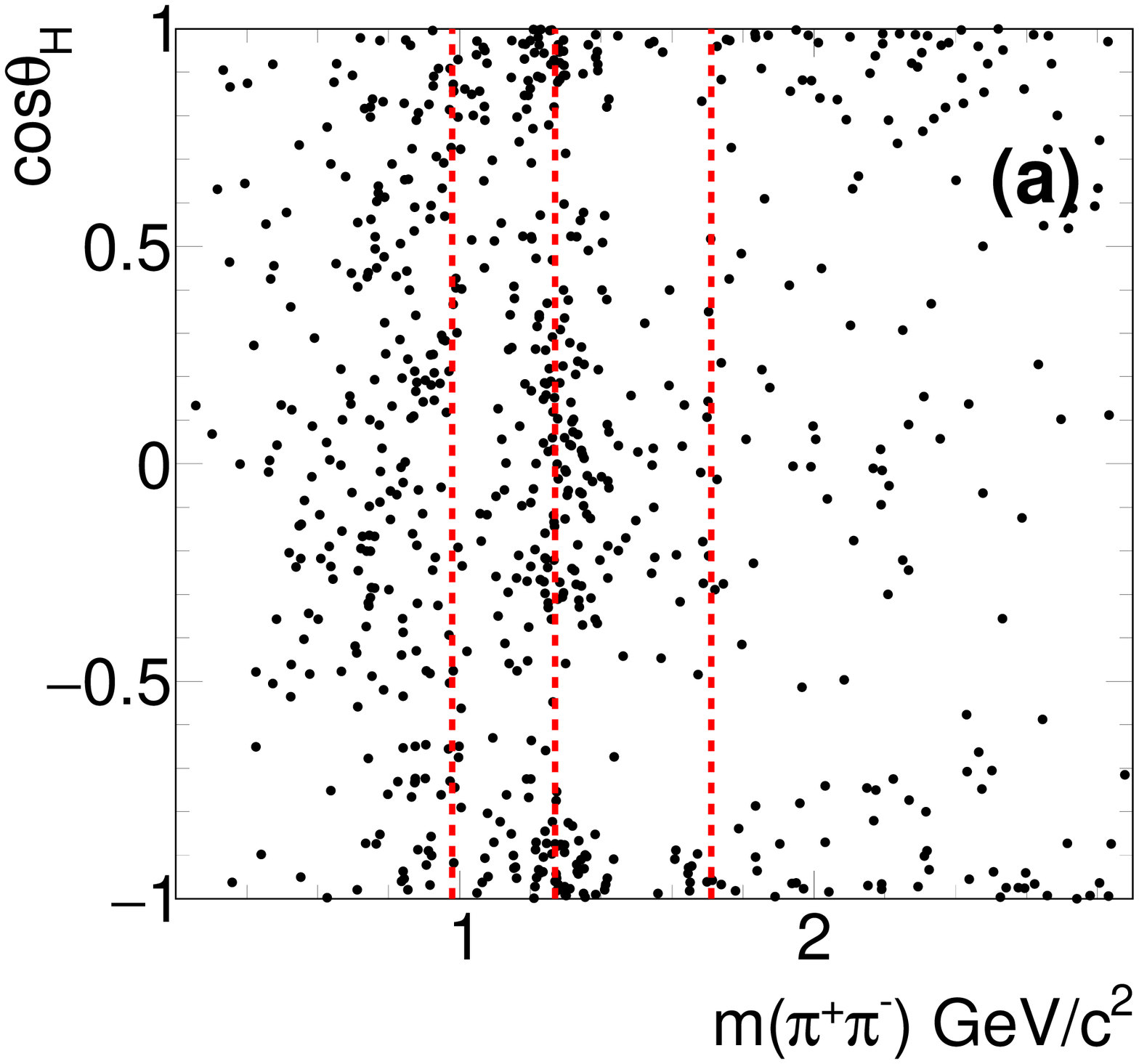}
\includegraphics[width=8.0cm]{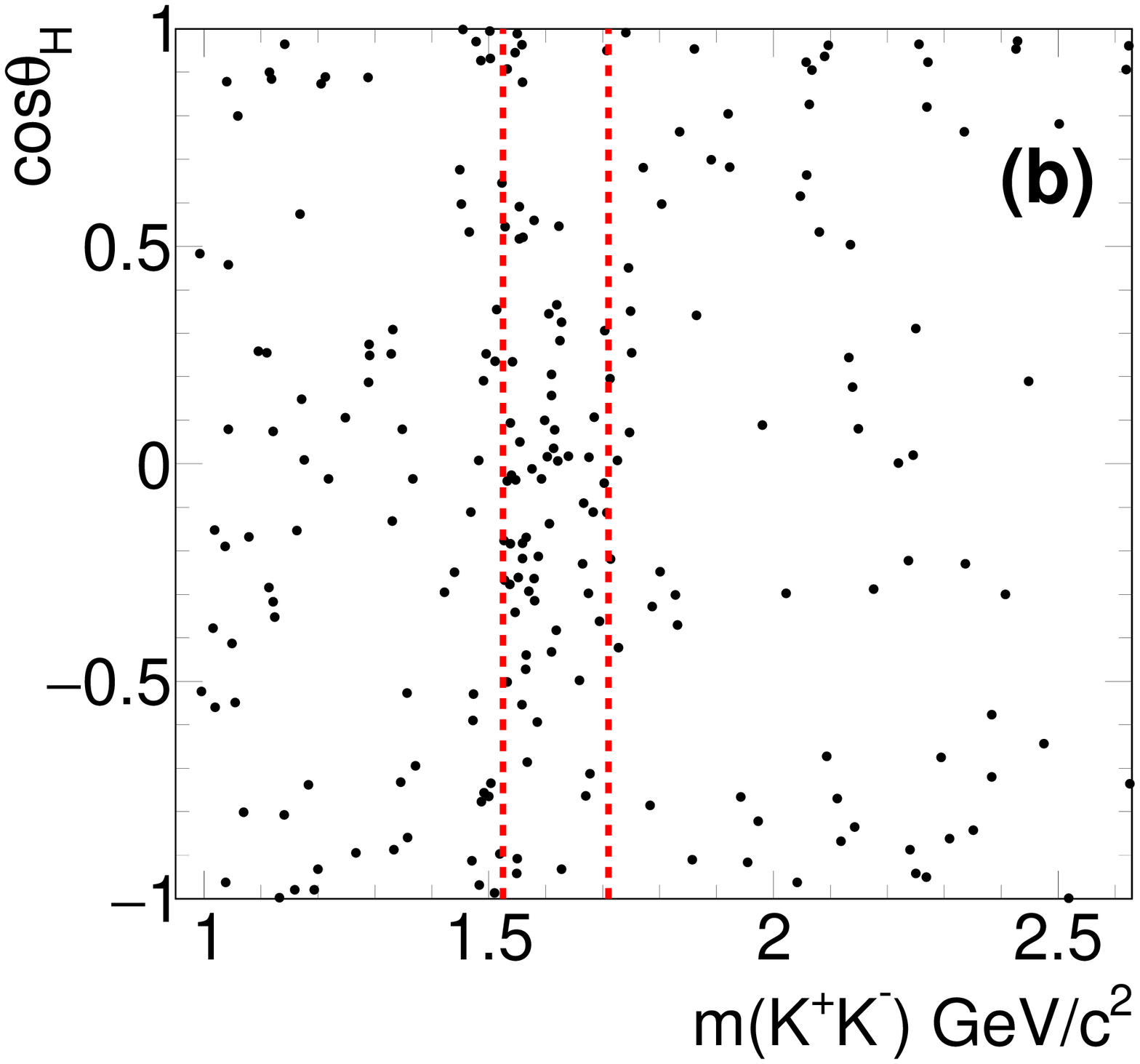}
\caption{(a) Uncorrected $\cos\theta_H$ vs. $m(\pip \pim)$
  distributions for the combined \TwoS and \ThreeS datasets.
The vertical dashed (red) lines indicate the positions of $f_0(980)$,
$f_2(1270)$, and $f_0(1710)$.
(b) $\cos\theta_H$ vs. $m(\Kp \Km)$ for the combined \TwoS and \ThreeS datasets. The vertical dashed (red) lines indicate the positions of the $f_2'(1525)$ and $f_0(1710)$.}
\label{fig:fig12}
\end{center}
\end{figure*}

To obtain information on the angular momentum structure of the $\pip \pim$ and
$\kp \km$ systems in $\OneS \to \gamma h^+ h^-$
we study the dependence of the $m(h^+ h^-)$ mass on the helicity angle
$\theta_H$. Figure~\ref{fig:fig12} shows the scatter plot
$\cos\theta_{H}$ vs $m(\pip \pim)$ for the combined \TwoS and \ThreeS
datasets. We observe the spin 2 structure of the $f_2(1270)$.

A better way to observe angular effects is to 
plot the $\pip \pim$ mass spectrum weighted by the Legendre polynomial
moments, corrected for efficiency. 
In a simplified environment, the moments are related to the spin 0 ($S$) and spin 2 ($D$) amplitudes by the equations~\cite{costa}:

\begin{figure*}[!htb]
\begin{center}
\includegraphics[width=14cm]{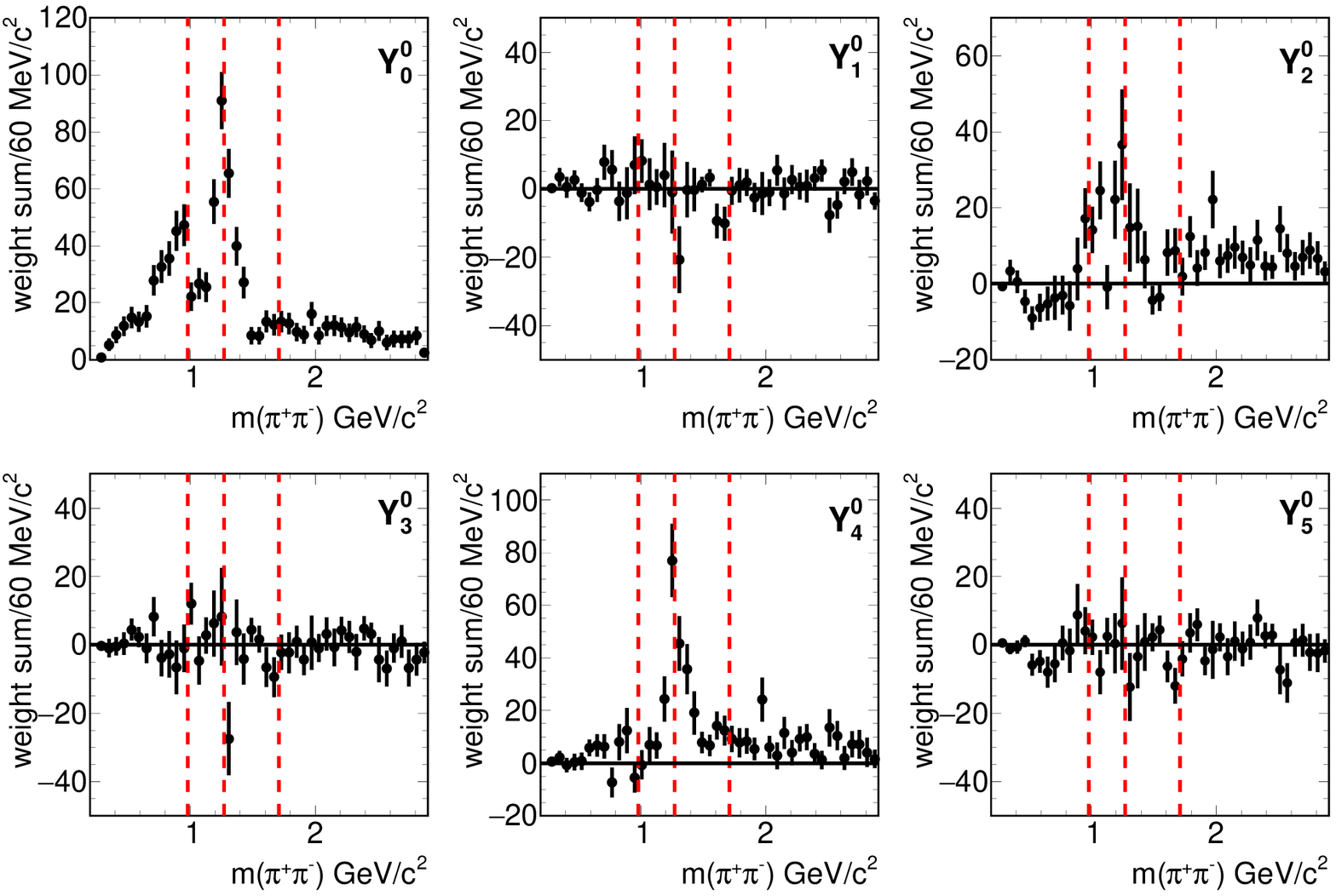}
\caption{The distributions of the unnormalized $Y^0_L$ moments for $\OneS \to
  \gamma \pip \pim$ as functions of the $\pip \pim$ mass corrected for
  efficiency. The lines indicate the positions of $f_0(980)$, $f_2(1270)$, and $f_0(1710)$.}
\label{fig:fig13}
\end{center}
\end{figure*}
\begin{figure*}[!htb]
\begin{center}
\includegraphics[width=14cm]{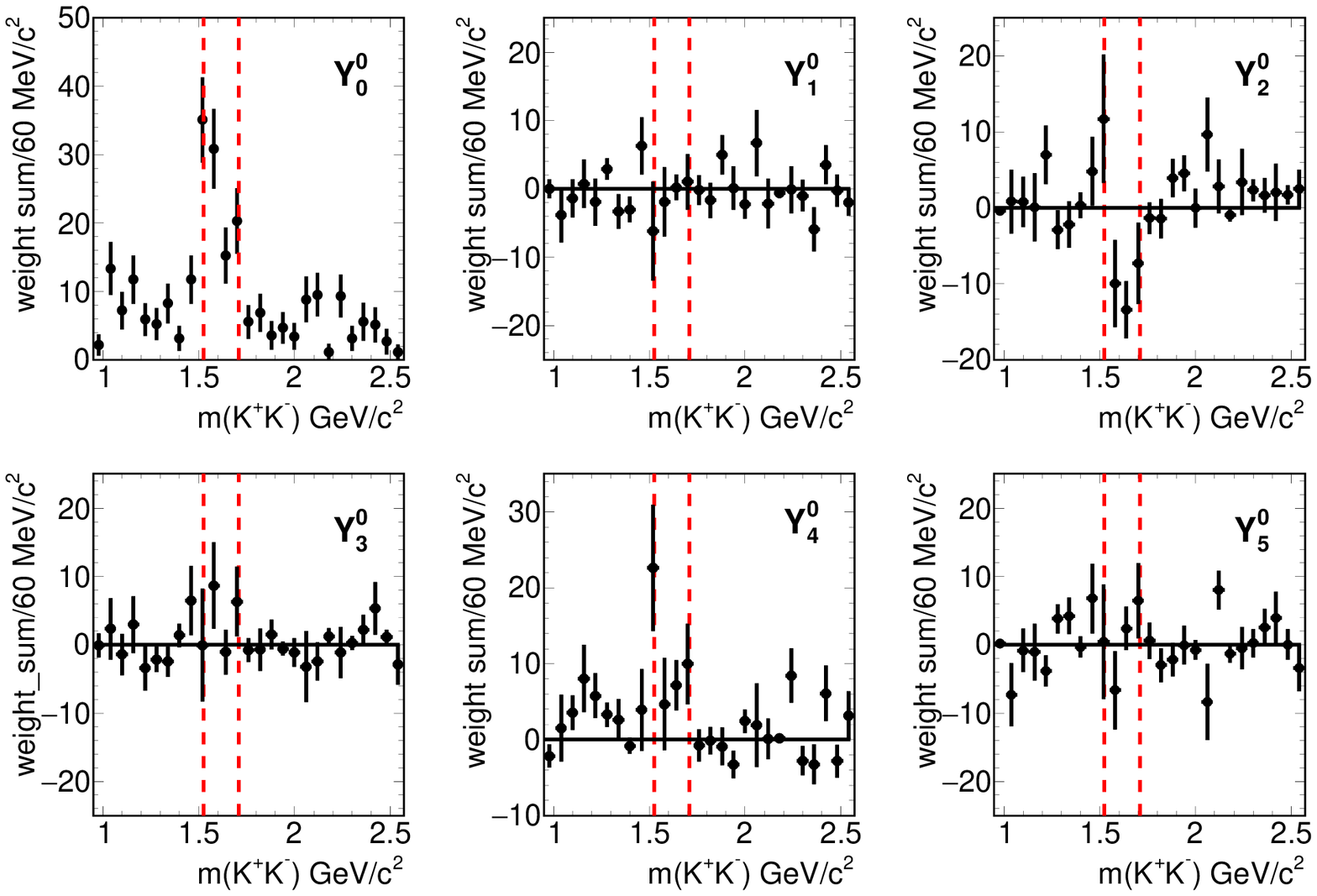}
\caption{The distributions of the unnormalized $Y^0_L$ moments for $\OneS \to
  \gamma \kp \km$ corrected for efficiency. The lines indicate the positions of the $f_2'(1525)$ and $f_0(1710)$.}
\label{fig:fig14}
\end{center}
\end{figure*}

\begin{equation}
\begin{split}
\sqrt{4 \pi}\langle Y_0^0\rangle=&S^2 + D^2,\\
\sqrt{4 \pi}\langle Y_2^0\rangle=&2SD\cos\phi_{SD} + 0.639 D^2,\\
\sqrt{4 \pi}\langle Y_4^0\rangle=&0.857D^2,\\
\end{split}
\label{eq:pwa}
\end{equation}

\noindent
where $\phi_{SD}$ is the relative phase.
Therefore we expect to observe spin 2 resonances in $\langle Y_4^0\rangle$ and
$S/D$ interference in $\langle Y_2^0\rangle$.
The results are shown in Fig.~\ref{fig:fig13}.
We clearly observe the $f_2(1270)$
resonance in $\langle Y_4^0\rangle$ and
a sharp drop in $\langle Y_2^0\rangle$ at the $f_2(1270)$ mass, indicating the
interference effect. 
The distribution of $\langle Y_0^0\rangle$ is just the scaled $\pip \pim$ mass distribution, corrected for efficiency.
Odd $L$ moments are sensitive to the $\cos\theta_{H}$ forward-backward asymmetry and show weak activity at the
position of the $f_2(1270)$ mass. Higher moments are all consistent with zero.

Similarly, we plot in Fig.~\ref{fig:fig14} the $\kp \km$ mass spectrum weighted by the Legendre polynomial moments, corrected for efficiency.
We observe signals of the $f_2'(1525)$ and $f_0(1710)$ in $\langle Y_4^0\rangle$ and activity due to $S/D$ interference effects in the $\langle Y_2^0\rangle$ moment.
Higher moments are all consistent with zero.

\begin{figure*}[!htb]
\begin{center}
\includegraphics[width=14.0cm]{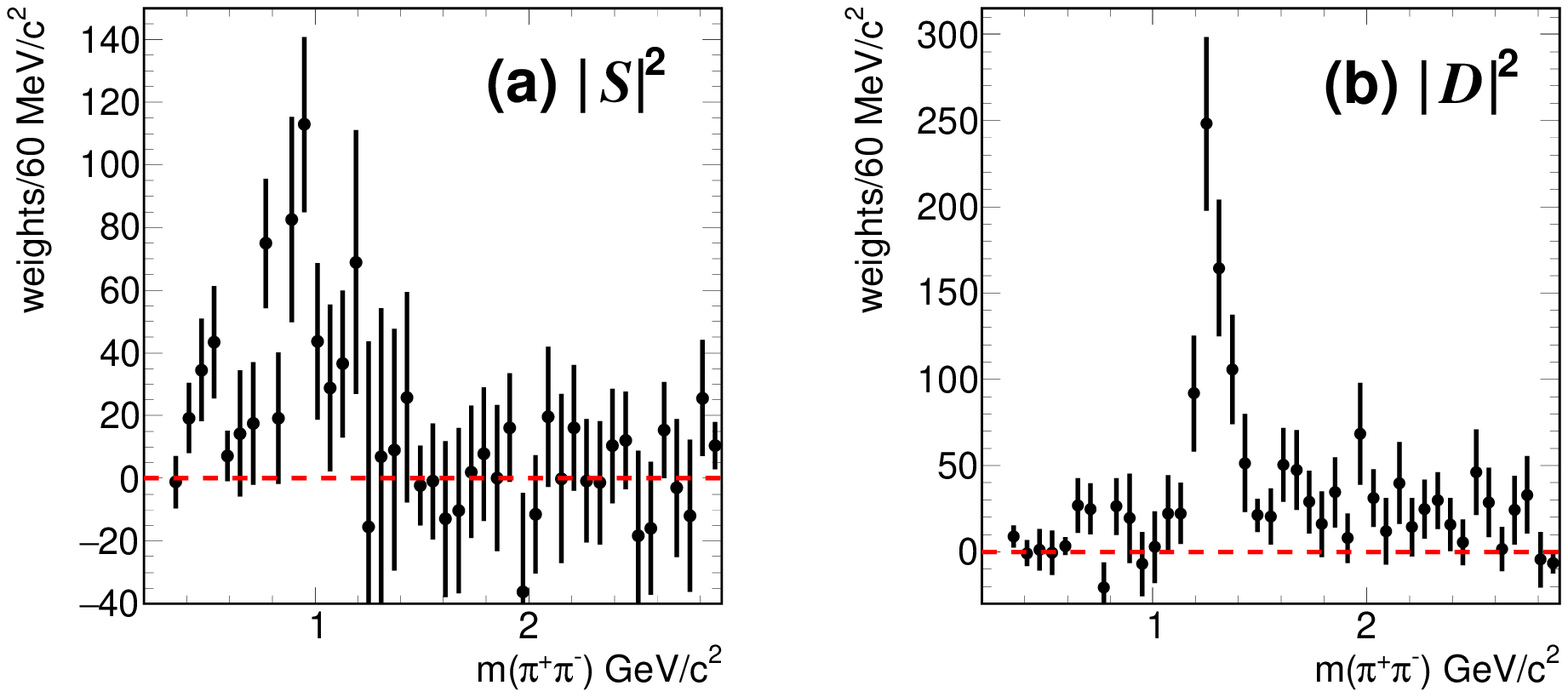}
\caption{Results from the simple PWA of the $\pip \pim$ mass spectrum for the \TwoS data. (a) $S$ and (b) $D$-wave contributions.} 
\label{fig:fig15}
\end{center}
\end{figure*}

\begin{figure*}[!htb]
\begin{center}
  \includegraphics[width=14.0cm]{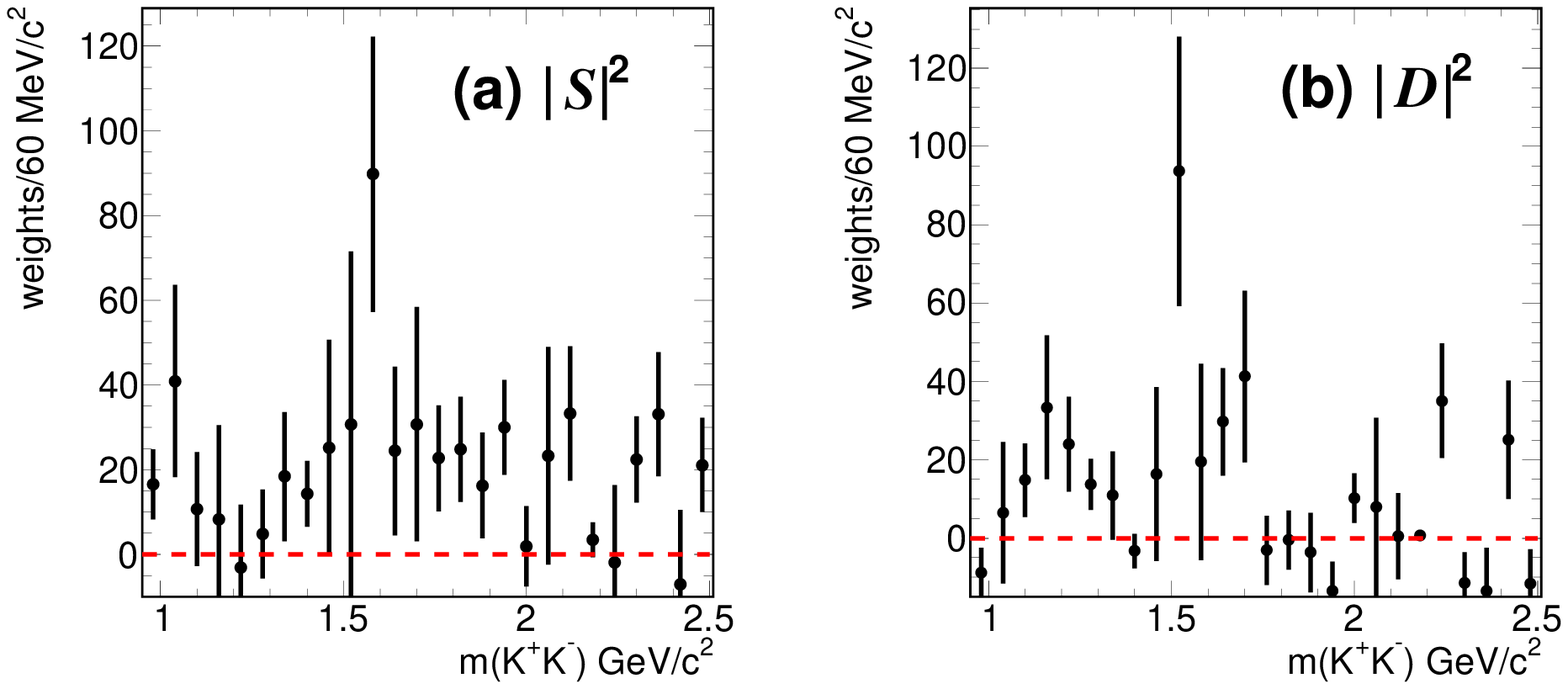}
\caption{Results from the simple PWA of the $K^+ K^-$ mass spectrum for the combined \TwoS/\ThreeS data. (a) $S$ and (b) $D$-wave contributions.} 
\label{fig:fig16}
\end{center}
\end{figure*}

Resonance angular distributions in radiative \OneS decays from \TwoS/\ThreeS decays are rather complex and will be studied in Sec.VIII.
In this section we perform a simplified Partial Wave Analysis (PWA) solving directly the system of Eq.~(\ref{eq:pwa}).
Figure~\ref{fig:fig15} and Fig.~\ref{fig:fig16} show the resulting $S$-wave and $D$-wave contributions to the $\pip \pim$ and $\kp \km$
mass spectra, respectively. Due to the presence of background in the threshold region, the $\pip \pim$ analysis is performed only on the \TwoS data.
The relative $\phi_{SD}$ phase is not plotted because it is affected by very large statistical errors.

We note that in the case of the $\pip \pim$ mass spectrum we obtain a good separation between $S$ and $D$-waves, with the presence
of an $f_0(980)$ resonance on top of a broad $f_0(500)$ resonance in the $S$-wave and a clean $f_2(1270)$ in the $D$-wave distribution.
Integrating the $S$-wave amplitude from threshold up to a mass of 1.5 \gevcc, we obtain an integrated, efficiency corrected
yield
\begin{equation}
  N(S{\scriptstyle{-}}{\rm wave}) = 629 \pm 128.
\end{equation}
\noindent
in agreement with the results from the fit to the $\pip \pim$ mass spectrum (see Table~\ref{tab:br2}).
We also compute the fraction of $S$-wave contribution in the $f_2(1270)$ mass region defined in Table~\ref{tab:pwa_h} and obtain $f_S(\pi^+ \pi^-) = 0.16 \pm  0.02$.

In the case of the $K^+ K^-$ PWA the structure peaking around 1500 \mevcc\ appears in both $S$ and $D$-waves suggesting the presence
of $f_0(1500)$ and $f_2'(1525)$. In the $f_0(1710)$ mass region there is not enough data to discriminate between the two different
spin assignments. This pattern is similar to that observed in the Dalitz plot analysis of charmless $B \to 3K$ decays~\cite{fx}.
Integrating the $S$ and $D$-wave contributions in the $f_2'(1525)/f_0(1500)$ mass region in the range given in Table~\ref{tab:pwa_h}, we obtain a fraction of $S$-wave contribution $f_S(K^+ K^-) = 0.53 \pm  0.10$.

\section{Spin-parity analysis}

\begin{figure}[!htb]
\begin{center}
\includegraphics[width=7cm]{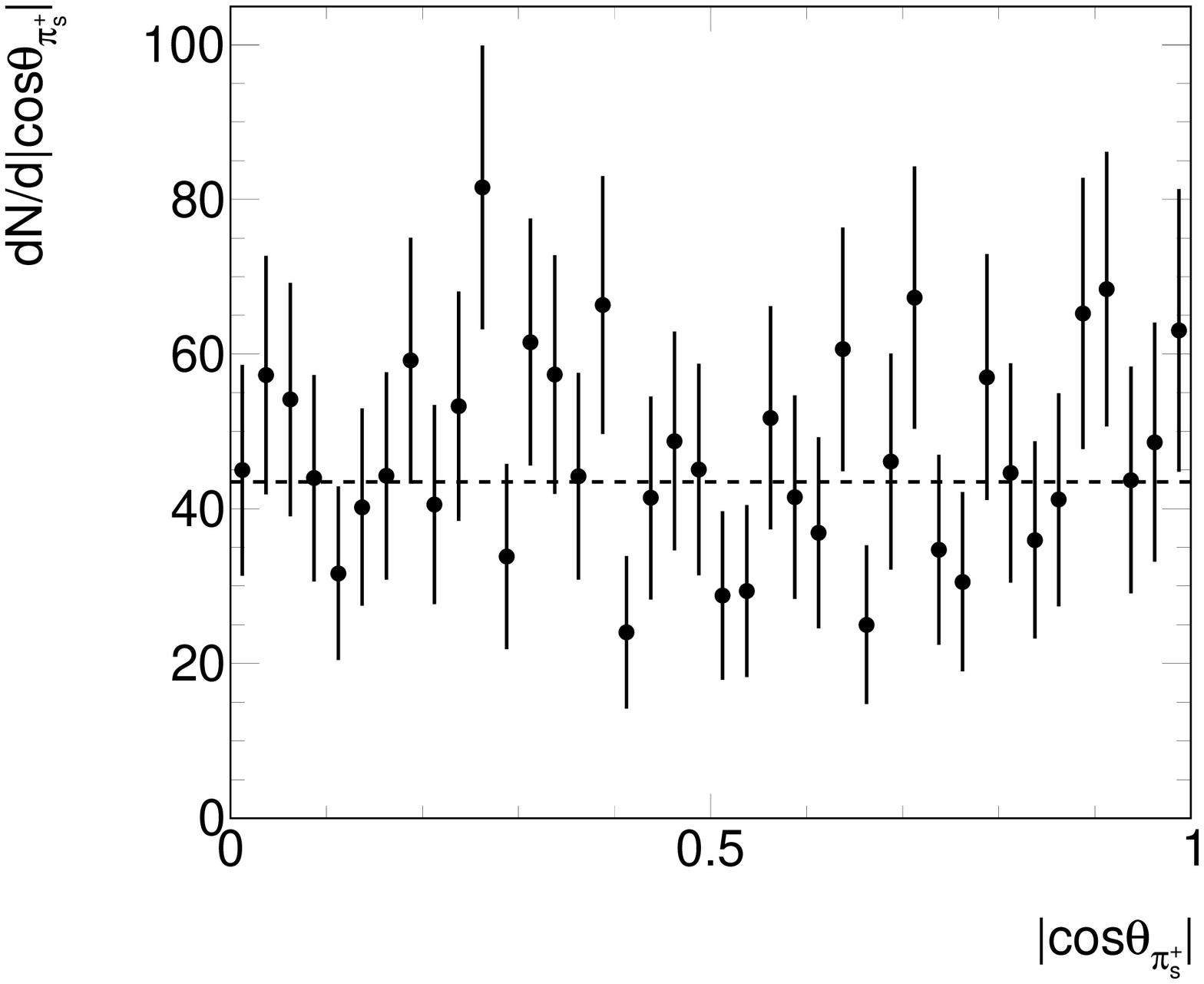}
\caption{Efficiency-corrected distribution of $\theta_{\pi}$ in
  the \TwoS data. The dashed line is the result of 
a fit to a uniform distribution.}
\label{fig:fig17}
\end{center}
\end{figure}

We compute the helicity angle $\theta_{\pi}$ defined as the angle formed by the $\pi^+_s$, in the $\pi^+_s \pi^-_s$ rest frame, with respect to the
direction of the $\pi^+_s \pi^-_s$ system in the $\OneS \pi^+_s \pi^-_s$ rest frame.
This distribution is shown in Fig.~\ref{fig:fig17}  for the \TwoS data and $\OneS \to \gamma \pip
\pim$, and is expected to be uniform
if $\pi^+_s \pi^-_s$ is an $S$-wave system.
The distribution is consistent with this hypothesis with a $p$-value of 65\%.

The $\Upsilon(nS)$ angular distributions are expressed in terms of $\theta_{\gamma}$ and $\theta_H$. 
Due to the decay chain used to isolate the \OneS radiative decays (see Eq.~(\ref{eq:twoa}) and Eq.~(\ref{eq:twob})), 
the \OneS can be produced with helicity 0 or 1 and the corresponding
amplitudes are labeled as $A_{00}$ and $A_{01}$, respectively. 
A spin 2 resonance, on the other hand, can have three helicity states, described by amplitudes $C_{10}$, $C_{11}$, and $C_{12}$.
We make use of the helicity formalism~\cite{cleothesis,richman} to derive
the angular distribution for a spin 2 resonance:

\begin{widetext}
\begin{eqnarray}
W_2(\theta_{\gamma},\theta_H) = \frac{dU(\theta_{\gamma},\theta_H)}{d\cos{\theta_\gamma}\, d\cos{\theta_H}}&=&\frac{15}{1024} |E_{00}|^2 \left[6 |A_{01}|^2 \left(22 |C_{10}|^2 + 8 |C_{11}|^2 + 9 |C_{12}|^2\right) \right.+\nonumber\\ 
    && 2 |A_{00}|^2 \left(22 |C_{10}|^2 + 24 |C_{11}|^2 + 9 |C_{12}|^2\right) + \nonumber\\
    && 24 \left(|A_{00}|^2 + 3 |A_{01}|^2\right) \left(2 |C_{10}|^2 - |C_{12}|^2\right) \cos{2 \theta_H}+\nonumber\\
    && 6 \left(|A_{00}|^2 \left(6 |C_{10}|^2 - 8 |C_{11}|^2 + |C_{12}|^2\right) + \right. \nonumber\\
    && \left. |A_{01}|^2 \left(18 |C_{10}|^2 - 8 |C_{11}|^2 +3 |C_{12}|^2\right)\right) \cos{4 \theta_H} -\nonumber\\ 
   && 2 \left(|A_{00}|^2 - |A_{01}|^2\right) \cos{2 \theta_\gamma} \left(22 |C_{10}|^2 - 24 |C_{11}|^2 + 9 |C_{12}|^2 + \right.\nonumber\\
   && 12 \left(2 |C_{10}|^2 - |C_{12}|^2\right) \cos{2 \theta_H} +\nonumber\\ 
    && \left.\left. 3 \left(6 |C_{10}|^2 + 8 |C_{11}|^2 +
       |C_{12}|^2\right) \cos{4 \theta_H}\right)\right].
\label{eq:spin2}
\end{eqnarray}
\end{widetext}

Ignoring the normalization factor $|E_{00}|^2$, there are two amplitudes describing the \OneS helicity states, which
can be reduced to one free parameter by taking the ratio
$|A_{01}|^2/|A_{00}|^2$. Similarly, the three amplitudes describing
the spin 2
helicity states, can be reduced to two free parameters by taking the
ratios $|C_{11}|^2/|C_{10}|^2$ and $|C_{12}|^2/|C_{10}|^2$. We
therefore have a total of three free parameters.

The expected angular distribution for a spin 0 resonance is given by
\begin{widetext}
\begin{equation}
W_0(\theta_{\gamma}) = \frac{dU({\theta_{\gamma})}}{d\cos\theta_{\gamma}} = \frac{3}{8}
|C_{10}|^2|E_{00}|^2 \left(|A_{00}|^2 + 3|A_{01}|^2 - (|A_{00}|^2 -
  |A_{01}|^2) \cos{2 \theta_\gamma} \right).
\label{eq:spin0}
\end{equation}
\end{widetext}
Ignoring the normalization factors $|C_{10}|^2$ and $|E_{00}|^2$, the distribution has only one free parameter, $|A_{01}|^2/|A_{00}|^2$.

We perform a 2D unbinned maximum likelihood fit for each resonance
region defined in Table~\ref{tab:pwa_h}. If $N$ is the number of available events, the likelihood function $\mathcal{L}$ is written as:
\begin{widetext}
  \begin{equation}
    \begin{split}
 \mathcal{L} = \prod_{n=1}^N \Big[ f_{\rm sig}\frac{\epsilon(\cos \theta_H,\cos
  \theta_{\gamma})W_s(\theta_H,\theta_{\gamma})}{\int W_s(\theta_H,\theta_{\gamma}) \epsilon(\cos \theta_H,\cos
  \theta_{\gamma})d\cos \theta_Hd\cos \theta_{\gamma}} + \\
      (1-f_{\rm sig})\frac{\epsilon(\cos \theta_H,\cos
        \theta_{\gamma})W_b(\theta_H,\theta_{\gamma})}{\int
        W_b(\theta_H,\theta_{\gamma}) \epsilon(\cos
        \theta_H,\cos \theta_{\gamma})d\cos \theta_Hd\cos
        \theta_{\gamma}}\Big]
 \end{split}
\end{equation}
\label{eq:lik}
\end{widetext}
where $f_{sig}$ is the signal fraction, $\epsilon(\cos \theta_H,\cos
\theta_{\gamma})$ is the fitted efficiency (Eq.~(\ref{eq:eff})), and
$W_s$ and $W_b$ are the functions describing signal and background contributions,
given by Eq.~(\ref{eq:spin2}) or Eq.~(\ref{eq:spin0}).
Since the background under the $\pip \pim$ and $\kp \km$
mass spectra is negligible in the low-mass regions, we include only
the tails of nearby adjacent resonances. 
In the description of the $\pip \pim$ data in the threshold region we
make use only of the \TwoS data because of the presence of a sizeable $\rho(770)^0$ background in the \ThreeS sample. 

\begin{figure}[!htb]
\begin{center}
\includegraphics[width=8.5cm]{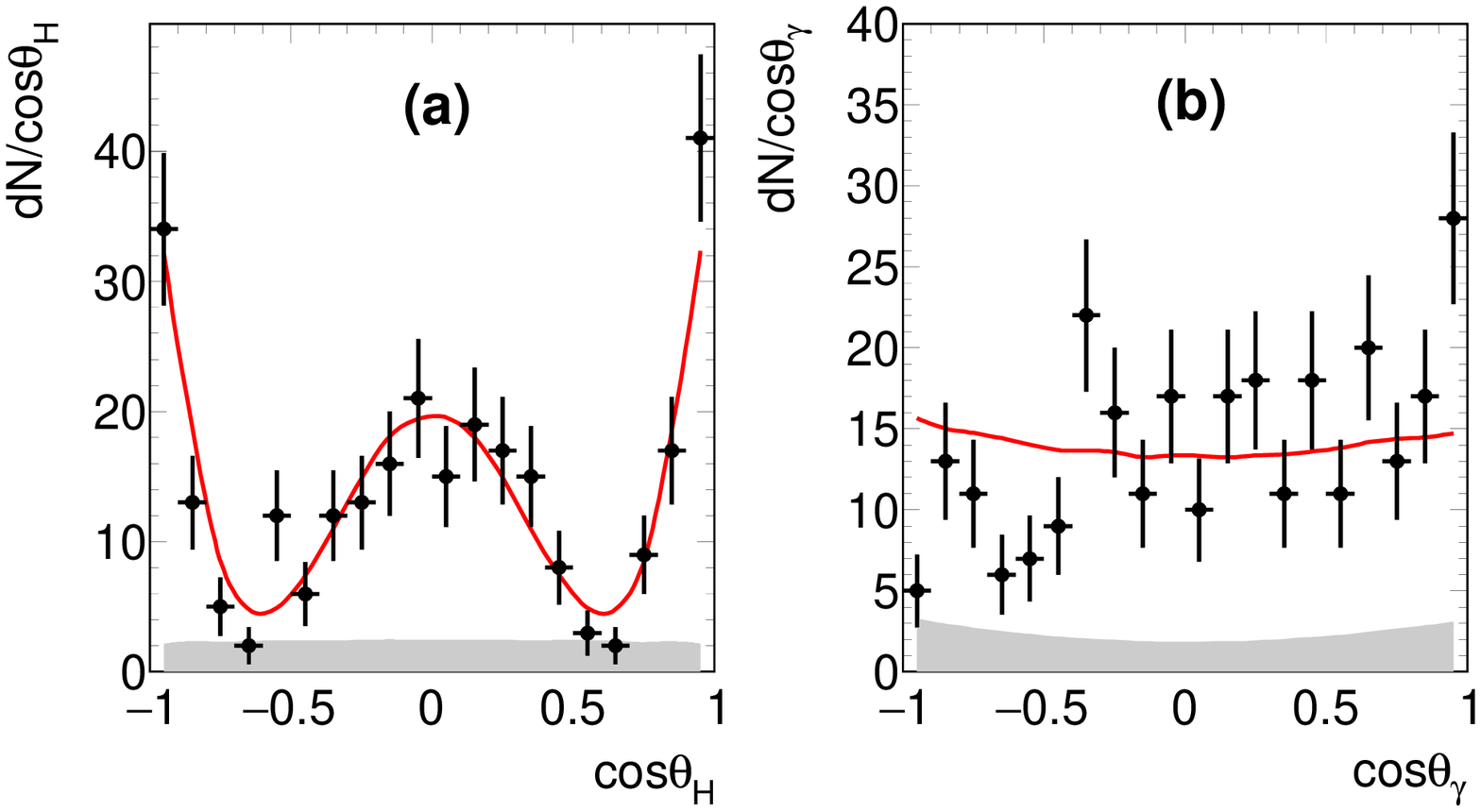}
\caption{Uncorrected
  (a) $\cos \theta_H$ and (b) $\cos \theta_{\gamma}$ distributions in the $f_2(1270) \to \pip \pim$ mass region.
  The full (red) lines are the projections from the fit
with the spin 2 hypothesis. The shaded (gray) area represents the $S$-wave background contribution.}
\label{fig:fig18}
\end{center}
\end{figure}

We first fit the $f_2(1270)$ angular distributions and allow a background contribution of 16\% (see Sect.~VII) from the $S$-wave having fixed
parameters.
Therefore an iterative procedure of fitting the $S$-wave and $f_2(1270)$ regions is performed.
Figure~\ref{fig:fig18} shows the uncorrected fit projections on $\cos
\theta_H$ and $\cos \theta_{\gamma}$. The $\cos \theta_{\gamma}$
spectrum is approximately uniform,
while $\cos\theta_H$ shows structures well-fitted by the spin 2 hypothesis. 
Table~\ref{tab:pwa_h} summarizes the results from the fits. We use as figures of merit $\chi_H=\chi^2(\cos\theta_H)$, $\chi_{\gamma}=\chi^2(\cos\theta_{\gamma})$ and their sum $\chi^2_t=(\chi_H+\chi_{\gamma})/{\rm ndf}$ computed as 
the $\chi^2$ values obtained from the $\cos\theta_H$ and $\cos\theta_{\gamma}$ projections, respectively.
We use ${\rm ndf}=N_{\rm cells}-N_{\rm par}$, where $N_{\rm par}$ is the number of free
parameters in the fit and $N_{\rm cells}$ is the sum of the number of bins
along the $\cos\theta_H$ and $\cos\theta_{\gamma}$ axes.
We note a good description of the $\cos\theta_H$ projection but a poor description of the $\cos\theta_{\gamma}$ projection.
This may be due to the possible presence of additional scalar components in the $f_2(1270)$ mass region, not taken into account in the formalism used in this analysis.

\begin{table*}[htb]
\begin{center}
\caption{Results from the helicity amplitude fits to resonances decaying to $\pip \pim$ and $\Kp \Km$.}
\label{tab:pwa_h}
\begin{tabular}{lccccccc}
\hline
\noalign{\vskip2pt}
Resonance & mass range (\gevcc) & events & spin & $\chi_H$, $\chi_{\gamma}$, $\chi^2_t/{\rm ndf}$ & $|A_{00}|^2/|A_{01}|^2$ &   &  \cr
\hline
\noalign{\vskip2pt}
$\pi \pi$ $S$-wave & 0.6-1.0 & 104 & 0 &5.8, 8.4, 14.3/19 & 0.09 $\pm$ 0.33 &   &       \cr
\hline
\noalign{\vskip2pt}
 &  &  & & & $|A_{01}|^2/|A_{00}|^2$ & $|C_{11}|^2/|C_{10}|^2$  & $|C_{12}|^2/|C_{10}|^2$ \cr
\hline
\noalign{\vskip2pt}
$f_2(1270) \to \pip \pim$ & 1.092-1.460 & 280 & 2 & 24.0, 46.0, 70/37 & 1.07 $\pm$ 0.31 & 0.00 $\pm$ 0.03  & 0.29 $\pm$ 0.08 \cr 
\hline
\noalign{\vskip2pt}
$f_2'(1525) \to \kp \km$ & 1.424-1.620 & 36  & 2 & 6.7, 1.8, 8.5/16 & $47.9 \pm 10.8$ & $0.42 \pm 0.36$ & $1.43 \pm 0.35$ \cr
$f_0(1500) \to \kp \km$ &             & 40  & 0 &               & $0.04 \pm 0.07$ &   &  \cr
\hline
\end{tabular}
\end{center}
\end{table*}

\begin{figure}[!htb]
\begin{center}
\includegraphics[width=8.5cm]{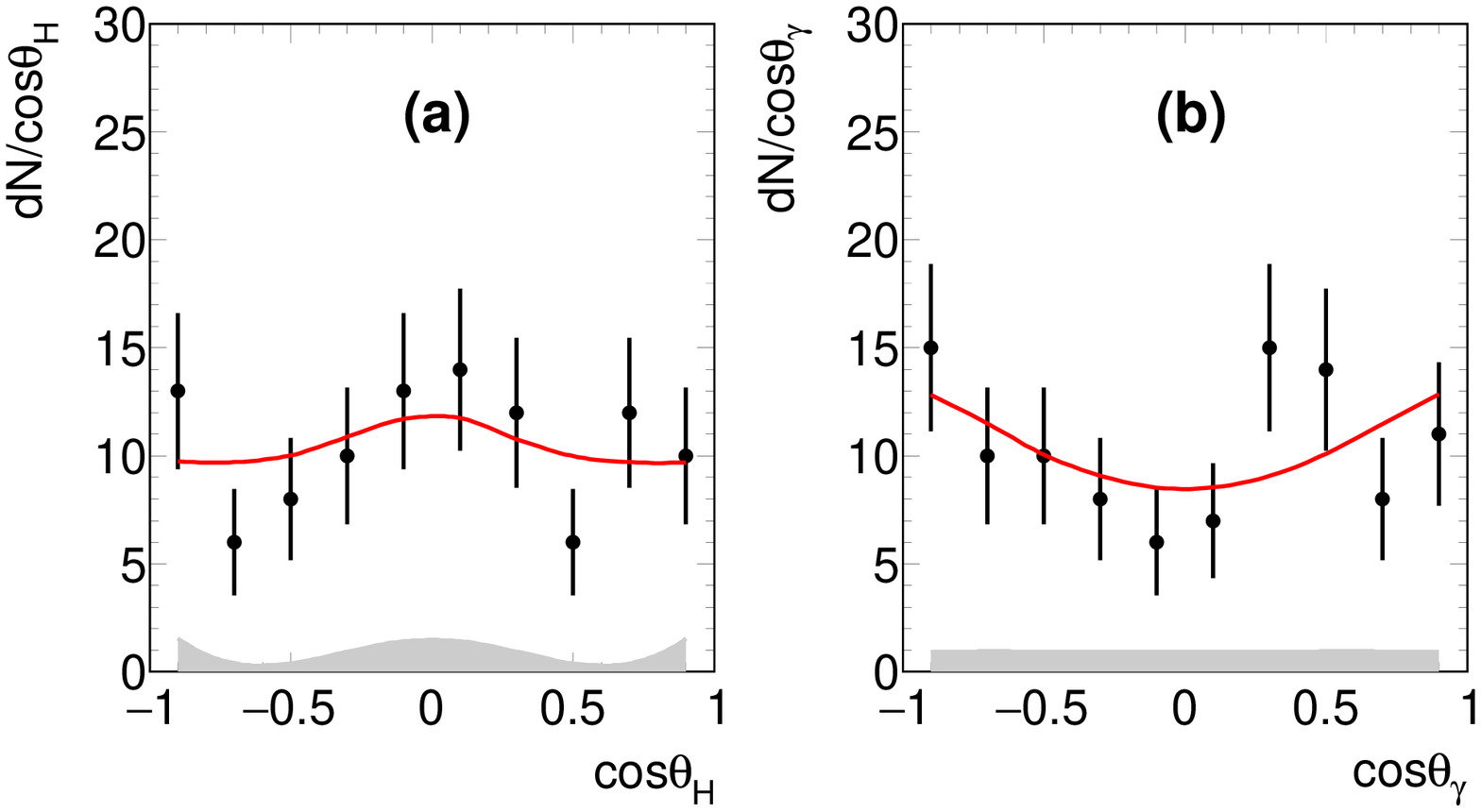}
\caption{Uncorrected
  (a) $\cos \theta_H$ and (b) $\cos \theta_{\gamma}$ distributions in the $S$-wave$\to \pip \pim$ mass region.
  The full (red) lines are the projections from the fit
using the spin 0 hypothesis. The shaded (gray) area represents the background contribution from 
the $f_2(1270)$.}
\label{fig:fig19}
\end{center}
\end{figure}

We fit the $S$-wave region in the $\pip \pim$ mass spectrum from the \TwoS
decay including as background the spin 2
contribution due to the tail of the $f_2(1270)$. The latter is estimated to contribute with a fraction
of 9\%, with parameters fixed to those obtained from the $f_2(1270)$
spin analysis described above. 
Figure~\ref{fig:fig19} shows the fit projections on the $\cos
\theta_H$ and $\cos \theta_{\gamma}$ distributions and Table~\ref{tab:pwa_h} gives details on the fitted parameters.
We obtain a good description of the data consistent with the spin 0 hypothesis.

\begin{figure}[!htb]
\begin{center}
\includegraphics[width=8.5cm]{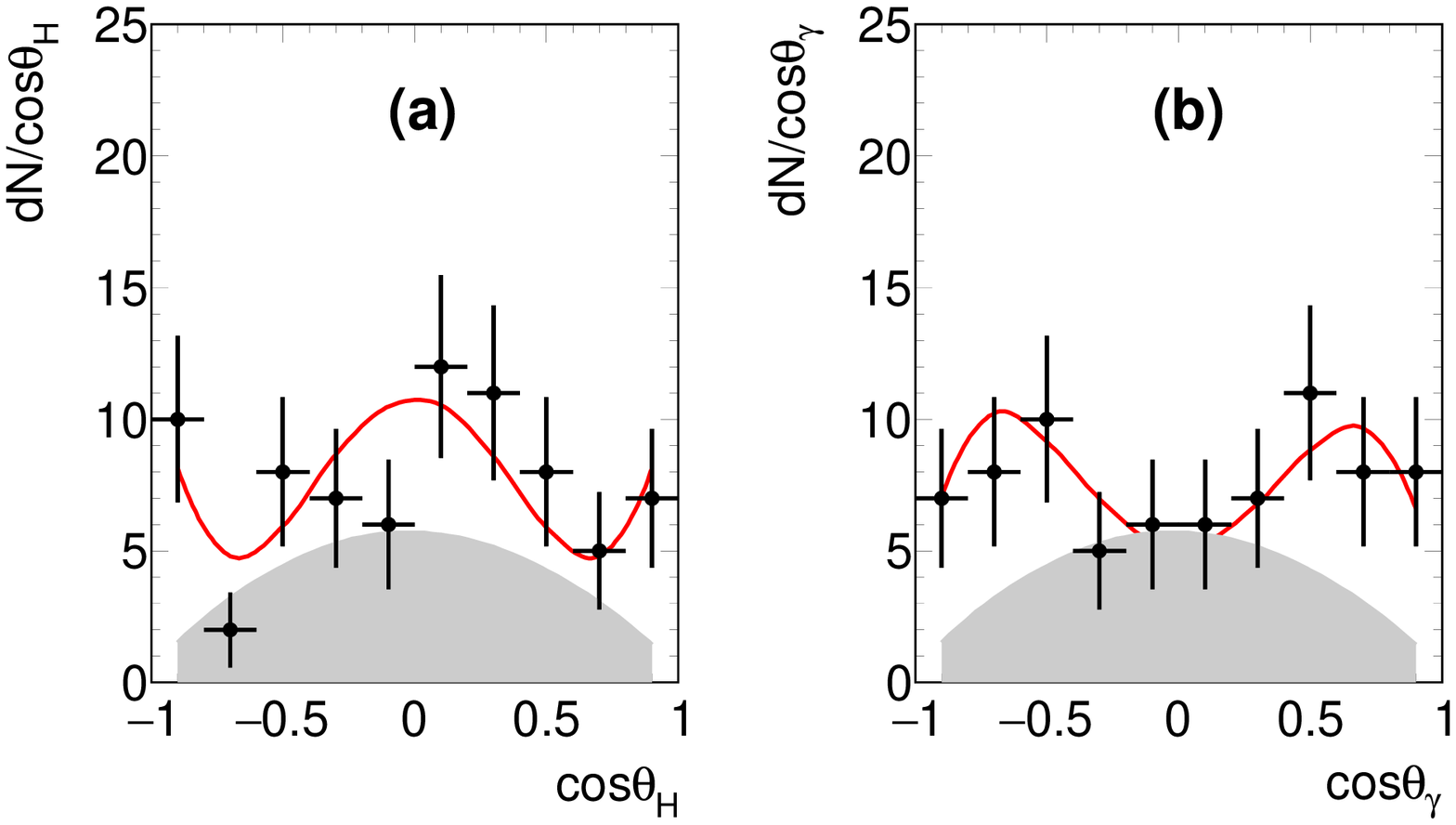}
\caption{Uncorrected
  (a) $\cos \theta_H$ and (b) $\cos \theta_{\gamma}$ distributions in the $f_J(1500)\to \kp \km$
  mass region. The full (red) lines are the projections from the fit using the superposition of spin-2 and spin-0 hypotheses.
  The shaded (gray) area represents the spin-0 contribution.}
\label{fig:fig20}
\end{center}
\end{figure}

We fit the $\kp \km$ data in the $f_J(1500)$ mass region, where many resonances can contribute:
$f_2'(1525)$, $f_0(1500)$~\cite{fx}, and $f_0(1710)$. We fit the data using a superposition of $S$ and $D$ waves,
having helicity contributions as free parameters, and free $S$-wave contribution.
We obtain an $S$-wave contribution of $f_S(K^+ K^-)=0.52 \pm 0.14$, in agreement with the estimate obtained in Sec.VI.
The helicity contributions are given in Table~\ref{tab:pwa_h} and fit projections are shown in Fig.~\ref{fig:fig20}, giving an adequate description of the data.
We assign the spin-2 contribution to the $f_2'(1525)$ and the spin-0 contribution to the $f_0(1500)$ resonance.
We also fit the data assuming the presence of the spin-2 $f_2'(1525)$ only hypothesis. We obtain a likelihood variation of $\Delta(-2\log \mathcal{L})=1.3$ for the
difference of two parameters between the two fits. Due the low statistics we cannot statistically distinguish between the two hypotheses.

\section{\boldmath Measurement of branching fractions}

We determine the branching fraction $\calB(R)$ for the decay of \OneS to photon and resonance $R$ using the expression

\begin{eqnarray}
  \calB(R) &=& \frac{N_R(\Upsilon(nS)\to \pi^+_s \pi^-_s \OneS(\to R \gamma))}{N(\Upsilon(nS)\to \pi^+_s \pi^-_s \OneS(\to \mup \mun))}\times \nonumber \\
&& \calB(\OneS \to \mup \mun),
\label{eq:br}
\end{eqnarray}
\noindent
where $N_R$ indicates the efficiency-corrected yield for the given
resonance. To reduce systematic
uncertainties, we first compute the relative branching fraction to the
reference channel
$\Upsilon(nS)\to \pip \pim \OneS(\to \mup \mun)$, which has the same
number of charged particles as the final states under study. We then
multiply the relative branching fraction by the well-measured branching
fraction $\calB(\OneS \to \mup \mun) =2.48 \pm 0.05$\%~\cite{PDG}.

We determine the reference channel corrected yield using the method of
``$B$-counting'', also used to obtain the number of produced \TwoS and \ThreeS~\cite{book}. Taking into account the known branching fractions of
$\TwoS/\ThreeS \to \pi^+_s \pi^-_s \OneS$ we obtain
\begin{equation}
  N(\TwoS \to \pi^+_s \pi^-_s \OneS (\to \mup \mun)) = (4.35 \pm 0.12_{\rm sys})\times 10^5
  \label{eq:bc1}
\end{equation}
and
\begin{equation}
  N(\ThreeS \to \pi^+_s \pi^-_s \OneS (\to \mup \mun)) = (1.32 \pm 0.04_{\rm sys}) \times 10^5
  \label{eq:bc2}
\end{equation}
\noindent events.
\noindent
As a cross-check we reconstruct $\Upsilon(nS)\to \pip \pim \OneS(\to \mup \mun)$ corrected for efficiency and obtain 
yields in good agreement with those obtained using the method of ``$B$-counting''.
 
Table~\ref{tab:brf} gives the measured branching fractions.
In all cases we correct the efficiency corrected yields for isospin and for
PDG measured branching fractions~\cite{PDG}.
In these measurements the $f_2(1270)$ yield is
corrected  first for the $\piz \piz$ (33.3\%) and
then for the $\pi \pi$  ($84.2^{+2.9}_{-0.9}\%$) branching fractions. 
We also correct the $\pi \pi$ $S$-wave and $f_0(1710)$ branching fractions for the $\piz \piz$ decay mode.
In the case of $f_J(1500) \to \kp \km$ the spin analysis reported in Sec.VI and Sec.VII gives indications
of the presence of overlapping $f_2'(1525)$ and $f_0(1500)$ contributions.
We give the branching fraction for $f_J(1500) \to \kp \km$ and, separately, for the $f_2'(1525)$ and $f_0(1500)$, 
where we make use of the $S$-wave contribution $f_S(K^+ K^-)=0.52 \pm 0.14$, obtained in Sec.VII.

The $f_2'(1525)$ branching fraction is corrected for the $K \bar K$ decay mode ($(88.7 \pm 2.2)\%$).
For all the resonances decaying to $K \bar K$, the branching fractions are corrected for the unseen $K^0\bar K^0$ decay mode (50\%).

For the $f_2(1270)$ and $f_0(1710)$ resonances decaying to $\pip
\pim$, the relative branching ratios are computed separately for the
\TwoS and \ThreeS datasets, obtaining good agreement.
The values reported in Table~\ref{tab:brf} are determined
using the weighted mean of the two measurements. 
\begin{table}[htb]
  \caption{Measured $\OneS \to \gamma R$ branching fractions.}
   \label{tab:brf}
\begin{center}
\begin{tabular}{lc}
  \hline
\noalign{\vskip2pt}
Resonance & \calB ($10^{-5}$) \cr
\hline
\noalign{\vskip2pt}
$\pi \pi$  $S$-wave & $\al \al \all 4.63 \pm 0.56 \pm 0.48$ \cr
$f_2(1270)$ & $\al\all 10.15 \pm 0.59 \al \all ^{+0.54}_{-0.43}$ \cr
$f_0(1710) \to \pi \pi$ & $\al \al \all 0.79 \pm 0.26 \pm 0.17$ \cr
\hline
$f_J(1500)\to K \bar K$ & $\al \al \all 3.97 \pm 0.52 \pm 0.55$ \cr
$f_2'(1525)$ & $\al \al \all 2.13 \pm 0.28 \pm 0.72$ \cr
$f_0(1500)\to K \bar K$ & $\al \al \all 2.08 \pm 0.27 \pm 0.65$ \cr
\hline
$f_0(1710) \to K \bar K $ & $\al \al \all 2.02 \pm 0.51 \pm 0.35$\cr
\hline
\end{tabular}
\end{center}
\end{table}

Since the reference channel has the same number of tracks as the final state, 
systematic uncertainties related to
tracking are negligible with respect to the errors due to other sources. 
The systematic uncertainty related to the ``$B$-counting'' estimate of the event yields in the denominator of Eq.~\ref{eq:br}
is propagated into the total systematic uncertainty on the branching fractions given in Table~\ref{tab:brf}.

Comparing with CLEO results, we note that our results on the  $S$-wave
contribution include the $f_0(980)$ and $f_0(500)$ contributions, while the CLEO
analysis determines the branching fraction for the peaking structure at the
$f_0(980)$ mass. In the same way a direct comparison for the $f_2'(1525)$ branching fraction is not possible
due to the $f_0(1500)$ contribution included in the present analysis.
The branching fraction for the
$f_2(1270)$ is in good agreement.

We report the first observation of $f_0(1710)$ in \OneS radiative decay with a significance of $5.7\sigma$, combining $\pip \pim$ and $\Kp \Km$ data.
To determine the branching ratio of the $f_0(1710)$ decays to $\pi \pi$ and $K \bar K$, we remove all the systematic uncertainties related to the reference
channels and of the $\gamma$ reconstruction. Labeling with $N$ the efficiency-corrected yields for the two $f_0(1710)$ decay modes,
we obtain
\begin{eqnarray}
  \frac{\calB(f_0(1710)\to \pi\pi)}{\calB(f_0(1710)\to K \bar K)} &=& \frac{N(f_0(1710) \to \pi \pi)}{N(f_0(1710) \to K \bar K)} \nonumber \\
  &=& 0.64 \pm 0.27_{\rm stat} \pm 0.18_{\rm sys},
\end{eqnarray}
in agreement with the world average value of $0.41^{+0.11}_{-0.17}$~\cite{PDG}.

\section{Summary}

We have studied the \OneS radiative decays to \gpipi and \gkk using data 
recorded with the \babar\ detector operating at the SLAC PEP-II
asymmetric-energy \epem\ collider at center-of-mass energies at the
\TwoS and \ThreeS resonances, using integrated luminosities of 13.6 \invfb and 28.0 \invfb, respectively.
The \OneS resonance is reconstructed from the decay chains
$\Upsilon(nS)\to \pi^+ \pi^- \OneS$, $n=2,3$.
Spin-parity analyses and branching fraction measurements are reported for the resonances 
observed in the \pipi and \kk mass spectra.
In particular, we report the observation of broad $S$-wave, $f_0(980)$, and $f_2(1270)$ resonances in the 
$\pip \pim$ mass spectrum. We observe a signal in the 1500 \mevcc\ mass region of the $\kp \km$ mass spectrum for which the
spin analysis indicates contributions from both $f_2'(1525)$ and $f_0(1500)$ resonances.
We also report observation of 
$f_0(1710)$ in both $\pip \pim$ and $\kp \km$ mass spectra with combined significance of $5.7\sigma$, and measure the relative branching fraction.
These results
may contribute to the long-standing issue of the identification of a scalar
glueball.

Reference~\cite{ochs} reports on a detailed discussion on the status of the search for the scalar glueball, listing as
candidates the broad $f_0(500)$, $f_0(1370)$, $f_0(1500)$, and $f_0(1710)$.
For this latter state, in the gluonium hypothesis, Ref.~\cite{zhu} computes a branching fraction of
$\calB(\OneS \to \gamma f_0(1710) = 0.96 ^{+0.55}_{-0.23} \times
10^{-4}$. Taking into account the presence of additional, not well
measured, $f_0(1710)$ decay modes, our result is consistent with this predicted branching fraction as well as with the dominance
of an $s \bar s$ decay mode.
For $f_0(1500) \to K \bar K$, ref.~\cite{he} expects a branching fraction $\calB(\OneS \to \gamma f_0(1500))$ in the range $2\sim 4 \times 10^{-5}$, consistent with our measurement.
The status of $f_0(1370)$ is controversial~\cite{pdg2006} as this state could just be an effect related to the broad $f_0(500)$.
Reference~\cite{zhu} estimates for $f_0(1370)$ a branching fraction of $\calB(\OneS \to \gamma f_0(1370))=3.2^{+1.8}_{-0.8} \times 10^{-5}$, in the range of our measurement
of the branching fraction of $\calB(\OneS \to \gamma (\pi \pi S\text{-wave}))$.

\section{Acknowledgments}

We are grateful for the 
extraordinary contributions of our \pep2\ colleagues in
achieving the excellent luminosity and machine conditions
that have made this work possible.
The success of this project also relies critically on the 
expertise and dedication of the computing organizations that 
support \babar.
The collaborating institutions wish to thank 
SLAC for its support and the kind hospitality extended to them. 
This work is supported by the
US Department of Energy
and National Science Foundation, the
Natural Sciences and Engineering Research Council (Canada),
the Commissariat \`a l'Energie Atomique and
Institut National de Physique Nucl\'eaire et de Physique des Particules
(France), the
Bundesministerium f\"ur Bildung und Forschung and
Deutsche Forschungsgemeinschaft
(Germany), the
Istituto Nazionale di Fisica Nucleare (Italy),
the Foundation for Fundamental Research on Matter (The Netherlands),
the Research Council of Norway, the
Ministry of Education and Science of the Russian Federation,
Ministerio de Economia y Competitividad (Spain), and the
Science and Technology Facilities Council (United Kingdom).
Individuals have received support from 
the Marie-Curie IEF program (European Union), the A. P. Sloan Foundation (USA) 
and the Binational Science Foundation (USA-Israel).
We acknowledge P. Colangelo for useful suggestions.

\renewcommand{\baselinestretch}{1}

\end{document}